\documentclass[aps,superscriptaddress,twocolumn,showpacs,dvips]{revtex4}
\usepackage{feynmp}
\usepackage{amsmath}
\usepackage{amssymb}
\usepackage{epsfig}
\usepackage{graphicx}
\usepackage{subfigure}
\usepackage{hyperref}
\usepackage{bbm}
\usepackage{color}

\newcommand{\Rmnum}[1]{\expandafter\@slowromancap\romannumeral #1@}
\allowdisplaybreaks[3]

\begin{document}

\title{Superconductivity around nematic quantum critical point in two-dimensional metals}

\author{Guo-Zhu Liu}
\altaffiliation{Corresponding author: gzliu@ustc.edu.cn}
\affiliation{Department of Modern Physics, University of Science and
Technology of China, Hefei, Anhui 230026, P. R. China}
\author{Jing-Rong Wang}
\affiliation{Anhui Province Key Laboratory of Condensed Matter
Physics at Extreme Conditions, High Magnetic Field Laboratory of the
Chinese Academy of Science, Hefei, Anhui 230031, P. R. China}

\begin{abstract}
We study the properties of $s$-wave superconductivity induced around
a nematic quantum critical point in two-dimensional metals. The
strong Landau damping and the Cooper pairing between incoherent
fermions have dramatic mutual influence on each other, and hence
should be treated on an equal footing. This problem is addressed by
analyzing the self-consistent Dyson-Schwinger equations for the
superconducting gap and Landau damping rate. We solve the equations
at zero temperature without making any linearization, and show that
the superconducting gap is maximized at the quantum critical point
and decreases rapidly as the system departs from this point. The
interplay between nematic fluctuation and an additional pairing
interaction, caused by phonon or other boson mode, is also
investigated. The total superconducting gap generated by such
interplay can be several times larger than the direct sum of the
gaps separately induced by these two pairing interactions. This
provides a promising way to achieve remarkable enhancement of
superconductivity.
\end{abstract}


\maketitle


\section{Introduction}

Conventional superconductors are well described by the
Bardeen-Cooper-Schieffer (BCS) theory \cite{BCS}, which gives a
microscopic mechanism for the superconducting (SC) transition and a
reliable quantitative estimate for the SC gap $\Delta$ and critical
temperature $T_c$ when the electron-phonon coupling is not strong.
BCS theory and its later extension by Eliashberg \cite{Eliashberg}
are firmly based on the validity of Fermi liquid (FL) theory
\cite{AGD} that is known to be perfectly applicable to normal
metals. When a net attraction is achieved, Cooper pairing is
realized, driving the SC phase transition.

The discovery of superconductivity in heavy fermion materials
\cite{Heavy}, cuprates \cite{Lee06, Fradkin15, Keimer15}, and iron
pnictides \cite{Chubukov12, Fernandes14, Shibauchi14, Kuo16,
Coldea17} has stimulated intensive research activities in the past
four decades. Many of these superconductors exhibit two salient
universal features: the existence of a SC dome with a maximal $T_c$;
the emergence of non-FL behavior in the non-SC phase. A great
challenge of condensed matter physics is to develop a unified
framework to account for these two features \cite{Heavy, Lee06,
Fradkin15, Keimer15, Chubukov12, Fernandes14, Shibauchi14, Kuo16}.
Based on existing experiments, it is broadly expected that the
observed NFL behavior and dome-shaped SC boundary might be caused by
certain quantum critical point (QCP). However, although this
scenario seems very promising, it proves very difficult to determine
the intrinsic correlation among the quantum criticality, NFL
behavior, and SC dome. Another unusual, but less universal, feature
is that, $T_c$ can be much higher in some cuprates and iron-based
superconductors than ordinary phonon-mediated superconductors. While
in principle Cooper pairing could be induced by several possible
bosonic modes, such as phonon, magnetic fluctuation, and nematic
fluctuation, none of them is able to account for all the basic
features observed in experiments.

In this paper, we study the fate of superconductivity formed in a
two-dimensional (2D) metal that is tuned close to a nematic quantum
phase transition \cite{Metlitski15, Raghu15, Raghu17, Scalapino15,
Einenkel15, Lederer15, Kivelson16, Kivelson17, Chubukov16, Mandal16,
Vishwanath16, LiZiXiang16, LiZiXiang17, Yamase13, Yamase16,
WangHuaJia17B, Labat17}. Such model system has wide applications
since electronic nematicity has already been observed in some
cuprates \cite{Fradkin15, Keimer15} and almost all iron-based
superconductors \cite{Chubukov12, Fernandes14, Shibauchi14, Kuo16,
Coldea17}. A special material is FeSe \cite{Coldea17, Bohmer17,
Hsu08, Song11}: it exhibits a clear nematic order; no magnetic order
is observed; the non-SC state is a NFL; $T_c$ is below $9$K but
significantly enhanced via K intercalation to reach $40$-$60$K;
there is a SC dome with $T_c$ being maximal near nematic QCP. It is
natural to expect that nematic order plays an essential role in the
formation of superconductivity in this material.

For a 2D metal, the quantum nematic fluctuation can lead to both
strong NFL behavior and Cooper pairing. There is a complicated
mutual influence between NFL behavior and Cooper pairing: strong
Landau damping shortens the fermion lifetime and accordingly may
affect the possibility of Cooper pairing; nonzero SC gap reduces the
space of final states into which fermions are scattered and hence
weakens the NFL behavior. It is hard to judge whether NFL behavior
favors superconductivity without doing concrete calculations. Since
the BCS-Eliashberg method become invalid in the NFL regime, it is
necessary to employ a generalized framework so that NFL behavior and
Cooper pairing can be treated on an equal footing.

\begin{figure}[htbp]
\center
\includegraphics[width=2.6in]{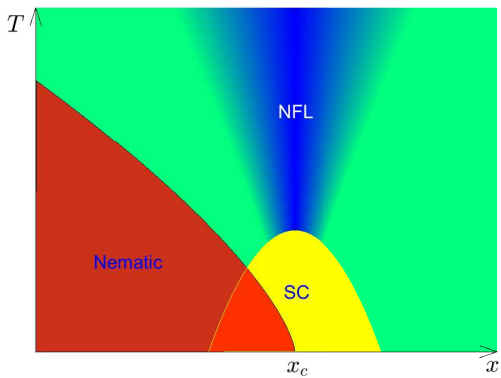}
\caption{Schematic phase diagram on the $x$-$T$ plane. We use $r = x
- x_c$ to denote the tuning parameter. The non-SC ground state is a
normal FL at large $r$, and a NFL at small $r$. SC gap is strongly
peaked at QCP. \label{Fig:phasediagram}}
\end{figure}

We will address the above issue by using the Dyson-Schwinger (DS)
integral equation approach, which is more general than
BCS-Eliashberg method. We first construct a set of self-consistently
coupled DS equations for the wave renormalization function and the
SC gap, and then solve them in an unbiased way. We are especially
interested in the magnitude of the SC gap, which could be measured
by experiments, including angle-resolved photoemission spectroscopy
and scanning tunneling microscope. For this purpose, we do not make
the linearizing approximation, which is valid only when $T \approx
T_c$, but directly solve the nonlinear DS equations. Based on the
solutions, the damping rate, SC gap, and their mutual influence can
be simultaneously determined. The vertex correction is simply
neglected in BCS-Eliashberg treatment. In our DS equation
calculation, we introduce a suitable \emph{ansatz} of the vertex
correction, and demonstrate that it can lead to substantial
influence on the gap size. Moreover, we also incorporate the
feedback of SC gap on the effective nematic propagator. The gap
strongly suppresses the low-energy DOS and strengthens the nematic
fluctuation, which enhances superconductivity.

After performing extensive calculations, we find that, the SC gap is
strongly peaked at the nematic QCP and rapidly suppressed as the
system moves away from the QCP into the disordered phase. There is a
clear dome-shaped curve of the SC gap, which is schematically shown
in Fig.~\ref{Fig:phasediagram}. Near the QCP, the NFL behavior is
hidden at energy scales below the SC gap, but may show its existence
at intermediate energy scales. We thus obtain a quantitative
description of the complicated correlation among nematic QCP, NFL
behavior, SC dome, and high $T_c$, which might be applicable to
doped FeSe and some cuprates.

Another crucial issue is to obtain a high, substantially enhanced
$T_c$. This can be naturally realized if we could find an efficient
way to promote the SC gap $\Delta$. We will study the interplay
between the nematic fluctuation and an additional pairing
interaction, which might arise from phonon or other type of bosonic
mode. A remarkable result is that the total gap generated by this
interplay is several times larger than the direct sum of the gaps
induced by two pairing mechanisms separately. Consequently, the
corresponding $T_c$ would be much larger than that produced by one
single pairing mechanism.

The rest of the paper is structures as follows. The model is given
in Sec.~\ref{Sec:Model}, and the self-consistent DS equations are
derived in Sec.~\ref{Sec:DerivationEq}. The solutions of DS
equations are presented and systematically analyzed in
Sec.~\ref{Sec:NumResults}. The main results are summarized in
Sec.~\ref{Sec:Summary}.

\section{Model Hamiltonian \label{Sec:Model}}

We consider a 2D metal at the border of nematic transition. The
low-energy effective model for a 2D metal close to a nematic QCP is
given by \cite{Metlitski15, Raghu15, Raghu17, Oganesyan01,
Metzner03, DellAnna06, Rech06, Metlitski2010, Holder15}
\begin{eqnarray}
L_{\psi} &=& \psi_{+\alpha}^{\dag} \left(\partial_{\tau} -
iv\partial_{x} - \frac{1}{2m}\partial_{y}^{2}\right)\psi_{+\alpha}
\nonumber \\
&& + \psi_{-\alpha}^{\dag} \left(\partial_{\tau} + iv\partial_{x} -
\frac{1}{2m}\partial_{y}^{2}\right)\psi_{-\alpha},
\\
L_{\phi} &=& \frac{N}{2e^{2}} \left(\partial_{y}\phi\right)^{2}
+ \frac{N_{f}r}{2}\phi^{2},
\\
L_{\psi\phi} &=& \phi\left(\psi_{+\alpha}^{\dag}\psi_{+\alpha} +
\psi_{-\alpha}^{\dag}\psi_{-\alpha}\right),
\\
L_{4f}&=&-\lambda\psi^{\dag}\psi^{\dag}\psi\psi.
\end{eqnarray}
Here, $L_{\psi}$ and $L_{\phi}$ are the Lagrangian densities for the
fermion field $\psi$ and the Ising-type nematic order parameter
$\phi$, respectively. The Yukawa coupling between $\psi$ and $\phi$
is described by $L_{\psi\phi}$. There is an additional short-ranged
BCS coupling term, given by $L_{4f}$, which might be induced by
exchanging phonons. We use $v$ to denote the Fermi velocity and $m$
the fermion mass. The $+$ and $-$ signs appearing in the fermion
field $\psi_{+,-}$ stand for the fermion excitations around two
patches near $\pm \mathbf{k}_{F}$ \cite{Metlitski15, Raghu15,
Raghu17, Oganesyan01, Metzner03, DellAnna06, Rech06, Metlitski2010,
Holder15}. Moreover, $\alpha = 1,2,...,N$ represents the fermion
flavor. The physical fermion flavor is $N = 2$, corresponding to the
two spin components.

To make this paper self-contained, we first sketch the computation
of the polarization function. The free fermion propagator is given
by
\begin{eqnarray}
G_{s}(\omega,\mathbf{k}) = \frac{1}{-i\omega + \xi_{\mathbf{k}}^{s}},
\end{eqnarray}
where $\xi_{\mathbf{k}}^{s} = \frac{\mathbf{k}^2}{2m}$ with $m$
being the fermion mass. At the Fermi surface, $\xi_{\mathbf{k}}^{s}$
can be simplified to
\begin{eqnarray}
\xi_{\mathbf{k}}^{s}=svk_{x}+\frac{k_{y}^{2}}{2m},
\label{Eq:FermionDispersionA}
\end{eqnarray}
where $k_x$ is the tangential component of momentum and $k_y$ the
perpendicular component. To the leading order of perturbative
expansion, the polarization is defined as
\begin{eqnarray}
\Pi(\Omega,\mathbf{q}) &=& N\sum_{s=\pm1}\int
\frac{d\omega}{2\pi} \frac{d^2\mathbf{k}}{(2\pi)^{2}}
G_{s}(\omega,\mathbf{k})\nonumber \\
&& \times G_{s}\left(\omega+\Omega,\mathbf{k}+\mathbf{q}\right).
\end{eqnarray}
This integral has already been computed in previous works, and it is
well-known that \cite{Metlitski15, Raghu15, Raghu17, Oganesyan01,
Metzner03, DellAnna06, Rech06, Metlitski2010, Holder15}
\begin{eqnarray}
\Pi(\Omega,\mathbf{q}) = -\frac{N}{2\pi^{2}} \frac{m}{v|q_{y}|}2\pi
i\mathrm{sgn}(\Omega) i\Omega = N\gamma\frac{|\Omega|}{|q_{y}|},
\label{Eq:PolarizationA}
\end{eqnarray}
where $\gamma=\frac{m}{\pi v}$. The dressed propagator of nematic
field $\phi$ has the form
\begin{eqnarray}
D(\omega,\mathbf{q}) = \frac{1}{N\left(\frac{q_{y}^{2}}{e^{2}} +
\gamma\frac{|\Omega|}{|q_{y}|}+r\right)},  \label{Eq:BosonPropagatorA}
\end{eqnarray}
which is independent of $q_{x}$. We have introduced a tuning
parameter $r$ to measure the distance of the system from the nematic
QCP. Depending on the material, $r$ could be doping concentration,
pressure, or external field. The nematic QCP is defined as $r = 0$.
Here, we approach to the QCP from the disordered phase, with $r$
decreasing from certain finite value down to zero continuously. The
ordered phase of nematic transition is more complicated, and thus we
leave for future work.

\section{Self-consistent Dyson-Schwinger equations \label{Sec:DerivationEq}}

The Yukawa interaction between gapless fermions and nematic
fluctuation can give rise to both strong Landau damping and Cooper
pairing. To study this problem, it is most convenient to define a
standard Nambu spinor:
\begin{eqnarray}
\Psi(\omega,\mathbf{k})=\left(\begin{array}{c}
\psi_{\uparrow}(\omega,\mathbf{k})
\\
\psi_{\downarrow}^{\dag}(-\omega,-\mathbf{k})
\end{array}\right).
\end{eqnarray}
Its conjugate is
\begin{eqnarray}
\Psi^{\dag}(\omega,\mathbf{k})=\left(\begin{array}{cc}
\psi_{\uparrow}^{\dag}(\omega,\mathbf{k}) &
\psi_{\downarrow}(-\omega,-\mathbf{k})
\end{array}\right).
\end{eqnarray}
For an ordinary FL (good) metal, one can write down the following
mean-field Lagrangian
\begin{eqnarray}
\mathcal{L} = \Psi^{\dag}\left(\begin{array}{cc}
i\omega_{n}-\xi_{\mathbf{k}} & \Delta
\\
\Delta^{*} & i\omega_{n}+\xi_{\mathbf{k}}
\end{array}\right)\Psi, \label{Eq:Lagrangian}
\end{eqnarray}
where $\Delta \propto \langle \psi_{\uparrow}\psi_{\downarrow}\rangle$ is
an $s$-wave SC gap. It is then easy to have a normal and an
anomalous Green's function:
\begin{eqnarray}
\mathcal{G}(\omega_{n},\mathbf{k}) = \frac{i\omega_{n} +
\xi_{\mathbf{k}}}{\omega_{n}^{2}+\xi_{\mathbf{k}}^{2}+\Delta^{2}},
\\
\mathcal{F}(\omega_{n},\mathbf{k}) = \frac{\Delta}{\omega_{n}^{2} +
\xi_{\mathbf{k}}^{2}+\Delta^{2}}.
\end{eqnarray}
The corresponding SC gap equation has the form
\begin{eqnarray}
\Delta &= &\lambda T\sum_{\omega_{n}} \int\frac{d^{d}
\mathbf{k}}{(2\pi)^{d}} \mathcal{F}(\omega_{n},\mathbf{k}). \\
&=& \lambda T\sum_{\omega_{n}}\int
\frac{d^{d}\mathbf{k}}{(2\pi)^{d}}\frac{\Delta}{\omega_{n}^{2} +
\xi_{\mathbf{k}}^{2}+\Delta^{2}}, \label{Eq:GapOriginal}
\end{eqnarray}
with $\lambda$ being the strength parameter for the attractive
force. We here focus on the case of $d=2$ spatial dimension, but the
extension to $d = 3$ is straightforward. The SC gap can be easily
obtained by solving this equation.

When a 2D metal is tuned on the border of long-range nematic order,
it becomes a bad metal due to the coupling of gapless fermions with
quantum critical nematic fluctuation. As the tuning parameter $r$
grows, driving the system to depart from nematic QCP, the NFL
behavior is gradually weakened. At finite $r$, the system exhibits
normal FL behavior at energy scales well below $r$ and unusual NFL
behavior at energy scale well above $r$. When the energy scale
becomes sufficiently large, the quantum nematic fluctuation
disappears. Thus, for finite $r$, NFL behavior actually emerges
within an intermediate range of energy scale. For very large $r$,
the range for NFL behavior to show up becomes extremely narrow, and
thus can be neglected. In this case, the mean-field Hamiltonian
(\ref{Eq:Lagrangian}) is definitely no longer sufficient to capture
the complicated mutual influence between FL behavior, NFL behavior,
and Cooper pairing. However, it is possible to properly generalize
(\ref{Eq:Lagrangian}) so as to describe the SC pairing in both the
FL and NFL regimes. For this purpose, we write the normal and
anomalous Green's functions in the following generic forms:
\begin{eqnarray}
\mathcal{G}'(\omega_{n},\mathbf{k}) &=& \frac{A_{1}(\omega_{n},
\mathbf{k})\omega_{n} + A_{2}(\omega_{n},\mathbf{k})
\xi_{\mathbf{k}}}{\Xi(\omega_{n},\mathbf{k})},
\\
\mathcal{F}'(\omega_{n},\mathbf{k}) &=&
\frac{\Delta(\omega_{n},\mathbf{k})}{\Xi(\omega_{n},\mathbf{k})}.
\end{eqnarray}
where
\begin{eqnarray}
\Xi(\omega_{n},\mathbf{k})=A_{1}^{2}(\omega_{n},\mathbf{k})
\omega_{n}^{2}+A_{2}^{2}(\omega_{n},\mathbf{k})\xi_{\mathbf{k}}^{2} +
\Delta^{2}(\omega_{n},\mathbf{k}). \nonumber
\end{eqnarray}

Here, $A_{1,2}(\omega_{n},\mathbf{k})$ are two renormalization
functions. The interaction-induced Landau damping is encoded in the
wave renormalization function $A_{1}(\omega_{n},\mathbf{k})$. It is
interesting that $A_{1}(\omega_{n},\mathbf{k})$ exhibits distinct
behaviors in NFL, FL, and fully gapped SC phases. For a NFL,
$A_{1}(\omega_{n},\mathbf{k})$ increases rapidly as the energy is
lowered, and eventually diverges in the zero-energy limit. For a FL,
$A_{1}(\omega_{n},\mathbf{k})$ is convergent at low energies. When
superconductivity is induced in a NFL metal,
$A_{1}(\omega_{n},\mathbf{k})$ approaches a finite value at energies
below the SC gap, but can still display NFL-like behaviors in an
intermediate range of energies. The fermion mass renormalization can
be obtained from $A_{1}(\omega_{n},\mathbf{k})$ and
$A_{2}(\omega_{n},\mathbf{k})$. Apparently, now
$\mathcal{G}'(\omega_{n},\mathbf{k})$ and
$\mathcal{F}'(\omega_{n},\mathbf{k})$ contain three important
interaction-induced effects: strong Landau damping, mass
renormalization, and SC gap generation. The functions
$A_{1,2}(\omega_{n},\mathbf{k})$ and $\Delta(\omega_{n},\mathbf{k})$
satisfy the following self-consistent DS integral equations:
\begin{eqnarray}
A_{1}(\varepsilon_{n},\mathbf{p})\varepsilon_{n} &=& \varepsilon_{n}
+ \int_{\omega_{n},\mathbf{k}}A_{1}(\omega_{n},\mathbf{k})\omega_{n}
F(\varepsilon_{n},\mathbf{p},\omega_{n},\mathbf{k}),
\nonumber \\
A_{2}(\varepsilon_{n},\mathbf{p})\xi_{\mathbf{p}} &=&
\xi_{\mathbf{p}} + \int_{\omega_{n},\mathbf{k}}
A_{2}(\omega_{n},\mathbf{k})\xi_{\mathbf{k}}
F(\varepsilon_{n},\mathbf{p},\omega_{n},\mathbf{k}), \nonumber \\
\Delta(\varepsilon_{n},\mathbf{p}) &=& \int_{\omega_{n},\mathbf{k}}
\Delta\left(\omega_{n},\mathbf{k}\right)
F(\varepsilon_{n},\mathbf{p},\omega_{n},\mathbf{k}) \nonumber \\
&& + \lambda \int_{\omega_{n},\mathbf{k}}
\frac{\Delta(\omega_{n},\mathbf{k})}{\Xi(\Omega_{n},\mathbf{k})},
\end{eqnarray}
where $\int_{\omega_{n},\mathbf{k}}\equiv T\sum_{\omega_{n}}
\int\frac{d^2\mathbf{k}}{(2\pi)^{2}}$ and
\begin{eqnarray}
F(\varepsilon_{n},\mathbf{p},\omega_{n},\mathbf{k}) = \frac{\Gamma
\left(\varepsilon_{n},\mathbf{p};\omega_{n},\mathbf{k}\right)}{\Xi(\omega_{n},\mathbf{k})}
D(\varepsilon_{n}-\omega_{n},\mathbf{p}-\mathbf{k}).
\end{eqnarray}
From these equations, we can see that the nematic fluctuation
contributes to both $A_{1,2}(\omega_{n},\mathbf{k})$ and
$\Delta(\omega_{n},\mathbf{k})$, whereas the BCS attraction makes no
contribution to $A_{1,2}(\omega_{n},\mathbf{k})$. In traditional
BCS-Eliashberg scheme for strongly coupled superconductors, the
vertex corrections are unimportant due to Migdal theorem. The
validity of this theorem relies on the fact that the fermion mass is
much smaller than the lattice mass. In the present case, there is no
guarantee that the vertex correction to the Yukawa coupling is
unimportant. In order not to underestimate the importance of vertex
correction, we have introduced a vertex function
$\Gamma\left(\varepsilon_{n},\mathbf{p};\omega_{n},\mathbf{k}\right)$.
In the most generic case, the vertex
$\Gamma\left(\varepsilon,\mathbf{p};\omega,\mathbf{k}\right)$ also
satisfies an integral equation that couples consistently to those of
$A_{1,2}(\varepsilon,\mathbf{p})$ and
$\Delta(\varepsilon,\mathbf{p})$. This would make the analysis
practically impossible. We will alternatively introduce some
suitable \emph{Ansatz} for the vertex function.

The three DS equations are intimately coupled to each other,
reflecting the fact that Landau damping effect, fermion mass
renormalization, and Cooper pairing have significant mutual
influence. It is thus difficult to perform numerical evaluations. In
the following, we will employ several further approximations. The
first approximation is to take the zero temperature limit by
employing the replacement $T\sum_{\omega_{n}} \rightarrow
\int_{-\infty}^{+\infty}\frac{d\omega}{2\pi}$. Since the dressed
nematic propagator $D(\varepsilon,\mathbf{p})$ does not depend on
$q_{x}$, the functions $A_{1,2}(\omega,\mathbf{k})$ are independent
of $k_{x}$ and thus can be written as $A_{1,2}(\omega,k_{y})$.
Accordingly, the vertex function is expressed in the form
$\Gamma\left(\varepsilon_{n},p_{y};\omega_{n},k_{y}\right)$. Now the
above self-consistent equations can be simplified to
\begin{eqnarray}
A_{1}(\varepsilon,p_{y})\varepsilon &=& \varepsilon +
\int_{\omega,k_{x},k_{y}} A_{1}(\omega,k_{y})\omega
F(\varepsilon,p_{y},\omega,k_{y}), \nonumber \\
A_{2}(\varepsilon,p_{y})\xi_{\mathbf{p}} &=& \xi_{\mathbf{p}}
+\int_{\omega,k_{x},k_{y}}A_{2}(\omega,k_{y})vk_{x}
F(\varepsilon,p_{y},\omega,k_{y}), \nonumber \\
A_{3}(\varepsilon,p_{y})\Delta &=&
\int_{\omega,k_{x},k_{y}}\Delta(\omega,k_{y})
F(\varepsilon,p_{y},\omega,k_{y}) \nonumber \\
&& +\lambda\int_{\omega,k_{x},k_{y}}
\frac{\Delta(\omega,k_{y})}{\Xi(\omega,k_{y})}, \label{Eq:GeneralEqsZeroT}
\end{eqnarray}
where
\begin{eqnarray}
F(\varepsilon,p_{y},\omega,k_{y}) = \frac{\Gamma
\left(\varepsilon,p_{y};\omega,k_{y}\right)}{\Xi(\omega,k_{y})}
D(\varepsilon-\omega,p_{y}-k_{y}),\nonumber
\end{eqnarray}
and
\begin{eqnarray}
\Xi(\omega,k_{y}) = A_{1}^{2}(\omega,k_{y})\omega^{2} +
A_{2}^{2}(\omega,k_{y})v^{2}k_{x}^{2} + \Delta^{2}(\omega,k_{y}).
\nonumber
\end{eqnarray}
Here, we use the notation $\int_{\omega,k_{x},k_{y}}\equiv
\int\frac{d\omega}{2\pi}\frac{dk_{x}}{2\pi}\frac{dk_{y}}{2\pi}$. A
transformation $vk_{x} + \frac{1}{2m}k_{y}^{2} \rightarrow vk_{x}$
has been utilized. It is now easy to verify that
\begin{eqnarray}
A_{2}(\varepsilon,p_{y}) = 1.
\end{eqnarray}
After performing the integration of $k_{x}$, we obtain
\begin{eqnarray}
A_{1}(\varepsilon,p_{y})\varepsilon &=& \varepsilon + \frac{1}{2v}
\int\frac{d\omega}{2\pi}\frac{dk_{y}}{2\pi}A_{1}(\omega,k_{y})\omega
F_{1}(\varepsilon,p_{y},\omega,k_{y}),\label{Eq:A12DGeneral}
\nonumber \\
\Delta(\varepsilon,p_{y}) &=& \frac{1}{2v}\int\frac{d\omega}{2\pi}
\frac{dk_{y}}{2\pi}\Delta(\omega,k_{y})
F_{1}(\varepsilon,p_{y},\omega,k_{y}) \nonumber \\
&& +\frac{\lambda}{2v}\int \frac{d\omega}{2\pi}
\frac{dk_{y}}{2\pi}\frac{\Delta(\omega,k_{y})}{J_{1}(\omega,k_{y})},
\label{Eq:Gap2DGeneral}
\end{eqnarray}
where
\begin{eqnarray}
F_{1}(\varepsilon,p_{y},\omega,k_{y})&=&\frac{\Gamma
\left(\varepsilon,p_{y};\omega,k_{y}\right)}
{J_{1}(\omega,k_{y})}D(\varepsilon-\omega,p_{y}-k_{y}), \nonumber \\
J_{1}(\omega,k_{y})&=&\sqrt{A_{1}^{2}(\omega,k_{y})\omega^{2} +
\Delta^{2}(\omega,k_{y})}.\nonumber
\end{eqnarray}

The vertex function needs to be specified at this stage. The
simplest choice is to adopt the bare vertex, i.e.,
\begin{eqnarray}
\Gamma\left(\varepsilon,p_{y};\omega,k_{y}\right) = 1.
\end{eqnarray}
This approximation is widely used in the BCS-Eliashberg treatment of
superconducting pairing \cite{Raghu17, Lederer15}, but is apparently
oversimplified. Here, we choose to consider the following
\emph{Ansatz}:
\begin{eqnarray}
\Gamma\left(\varepsilon,p_{y};\omega,k_{y}\right) = \frac{1}{2}
\left[A_{1}(\varepsilon,p_{y})+A_{1}(\omega,k_{y})\right],
\end{eqnarray}
which is symmetric under the exchange of energy-momentum variables.

We further suppose that the dependence of $A_{1}$ and $\Delta$ on
component $p_{y}$ is weak, namely
\begin{eqnarray}
A_{1}(\varepsilon,p_{y}) \rightarrow A_{1}(\varepsilon),\qquad
\Delta(\varepsilon,p_{y})\rightarrow\Delta\left(\varepsilon\right).
\end{eqnarray}
Here, we use the Fermi momentum $k_{F}$ to serve as the cutoff for
$k_{y}$. Now, the integration over $k_{y}$ appearing in the first
term of the gap equation in Eq.~(\ref{Eq:Gap2DGeneral}) is
convergent and can be carried out directly. For the rest terms of
Eq.~(\ref{Eq:Gap2DGeneral}), we define $|k_{y}| =
(e^{2}\gamma|\varepsilon-\omega|)^{1/3}x$ and then find that
\begin{eqnarray}
A_{1}(\varepsilon)\varepsilon &=& \varepsilon +
\frac{g\omega_{c}^{1/3}}{N} \int_{0}^{\omega_{c}}d\omega
A_{1}(\omega)\omega \nonumber \\
&& \times \left[F_{3}(\varepsilon-\omega) -
F_{3}(\varepsilon+\omega)\right],
\label{Eq:A11DGerneral}  \\
\Delta(\varepsilon) &=& \frac{g\omega_{c}^{1/3}}{N}
\int_{0}^{\omega_{c}} d\omega\Delta(\omega)
\left[F_{3}(\varepsilon-\omega)+F_{3}(\varepsilon+\omega) \right]
\nonumber \\
&& + \lambda' \int_{0}^{\omega_{c}}d\omega
\frac{\Delta(\omega)}{J_{2}(\omega)},
\label{Eq:Delta1DGerneral}
\end{eqnarray}
where
\begin{eqnarray}
F_{3}(\varepsilon\pm\omega) &=&
\frac{\Gamma\left(\varepsilon;\omega\right)}{J_{2}(\omega)}
\frac{1}{|\varepsilon\pm\omega|^{1/3}}\frac{3\sqrt{3}}{2\pi}
\nonumber \\
&&\times \int_{0}^{+\infty}dx \frac{x}{x^{3}+1 +
\Big(\frac{e}{\gamma|\varepsilon\pm\omega|}\Big)^{2/3} rx},
\end{eqnarray}
with $J_{2}(\omega)=\sqrt{A_{1}^{2}(\omega)\omega^{2} +
\Delta^{2}(\omega)}$. We have used the relations $A_{1}(\omega) =
A_{1}(-\omega)$ and $\Delta(\omega)=\Delta(-\omega)$, and defined
new parameters
\begin{eqnarray}
\lambda' = \frac{\lambda k_{F}}{2\pi^{2}v},\qquad g =
\frac{e^{4/3}}{6\sqrt{3}\pi v \gamma^{1/3}\omega_{c}^{1/3}}.
\end{eqnarray}
The upper limit of $x$ is taken to be infinity, which is justified
because the integration over $x$ is free of divergence. From
Eqs.~(\ref{Eq:A11DGerneral}) and (\ref{Eq:Delta1DGerneral}), we can
see that $1/N$ can serve as an effective expanding parameter for the
fermion-nematic interaction.

The equations (\ref{Eq:A11DGerneral}) and (\ref{Eq:Delta1DGerneral})
are applicable to both the FL regime and NFL regime, and also can
capture the FL-to-NFL crossover tuned by changing the energy scale,
which allows us to examine the mutual influence between Landau
damping and Cooper pairing as the system is approaching the nematic
QCP. At exactly the nematic QCP, $r = 0$ and we have
\begin{eqnarray}
A_{1}(\varepsilon)\varepsilon &=& \varepsilon +
\frac{g\omega_{c}^{1/3}}{N}\int_{0}^{\omega_{c}}
d\omega A_{1}(\omega)\omega K_{-}(\varepsilon,\omega), \\
\Delta(\varepsilon) &=& \frac{g\omega_{c}^{1/3}}{N}
\int_{0}^{\omega_{c}}d\omega\Delta(\omega)K_{+}(\varepsilon,\omega)
\nonumber \\
&& +\lambda'\int_{0}^{\omega_{c}}d\omega
\frac{\Delta(\omega)}{J_{2}(\omega)},
\end{eqnarray}
where
\begin{eqnarray}
K_{\pm}(\varepsilon,\omega) =
\frac{\Gamma\left(\varepsilon;\omega\right)}{J_{2}(\omega)}
\left(\frac{1}{|\varepsilon-\omega|^{1/3}} -
\frac{1}{|\varepsilon+\omega|^{1/3}}\right).
\end{eqnarray}
By solving the two equations self-consistently, we can get the
Landau damping rate, from $A_1(\varepsilon)$, and SC gap.

Our DS equations are quite general, and contain all the essential
information for Landau damping and Cooper pairing. They should
recover the results obtained previously in limiting cases. If the
nematic QCP is removed, $A_1 \equiv 1$, the DS equations are simply
reduced to the well-known BCS gap equation. An opposite limit is
reached by taking $\Delta = 0$, corresponding to the non-SC ground
state. In this case, the quantum nematic fluctuation leads to strong
fermion damping effect. After neglecting the vertex correction, we
obtain
\begin{eqnarray}
A_{1}(\varepsilon)\varepsilon \approx \varepsilon +
\frac{6g\omega_{c}^{1/3}}{N} \varepsilon^{2/3}
\end{eqnarray}
in the low-energy region $\varepsilon \ll \omega_{c}$. This is a
typical NFL behavior, and well consistent with the results reported
previously \cite{Oganesyan01, Metzner03, DellAnna06, Rech06,
Metlitski2010, Holder15}.

\section{Solutions of coupled equations \label{Sec:NumResults}}

In the limiting case with $\Delta = 0$, the system exhibits
well-known NFL behavior $\left(A_1 - 1\right)\varepsilon \propto
\varepsilon^{2/3}$ at $r=0$. At nonzero $r$, there is a crossover from
NFL regime to nomal FL regime as $r$ or the energy scale is varied.
Cooper pairing is formed on the basis of the NFL metal near nematic
QCP.

\begin{figure}[htbp]
\includegraphics[width=1.65in]{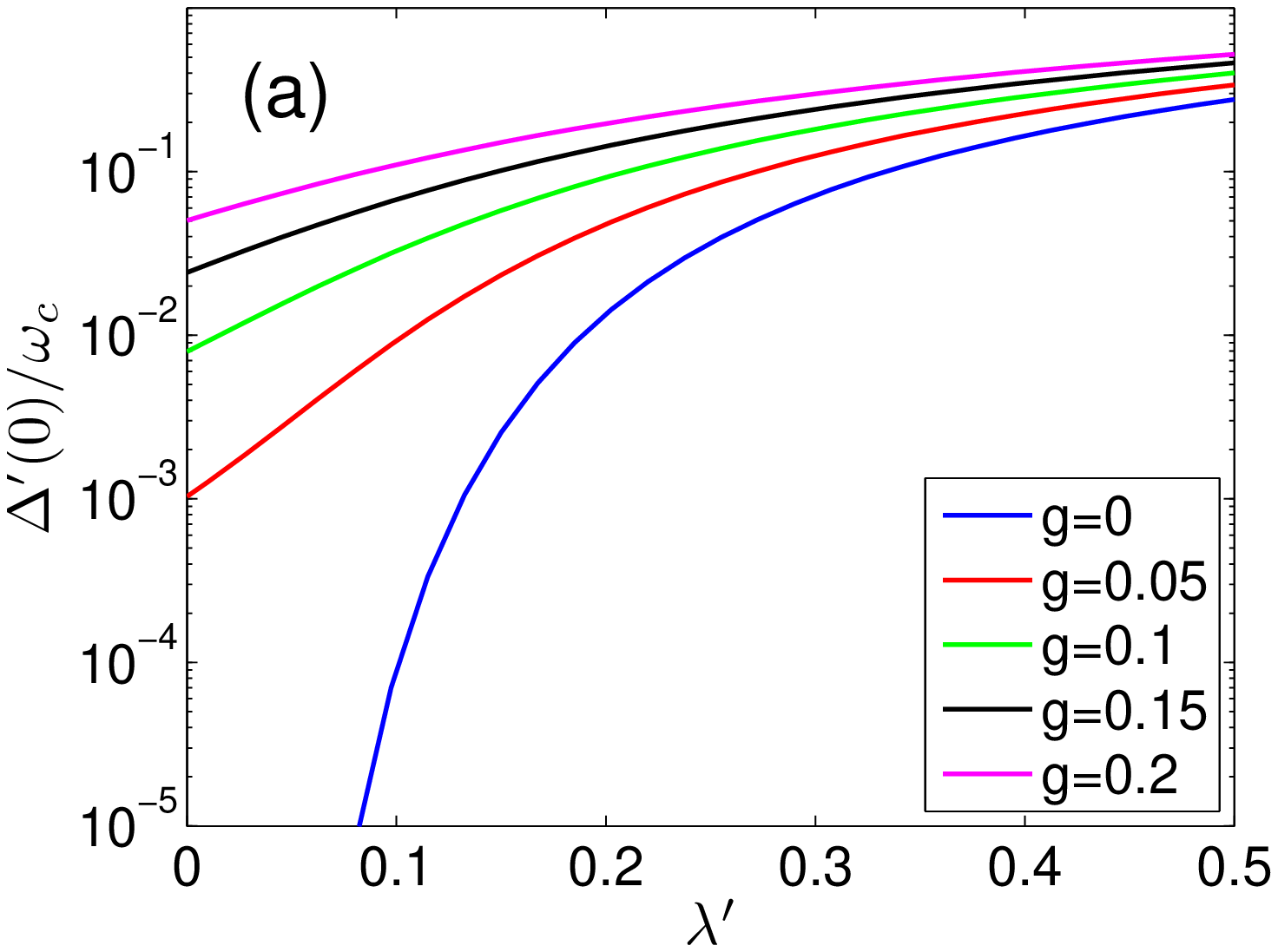}
\includegraphics[width=1.65in]{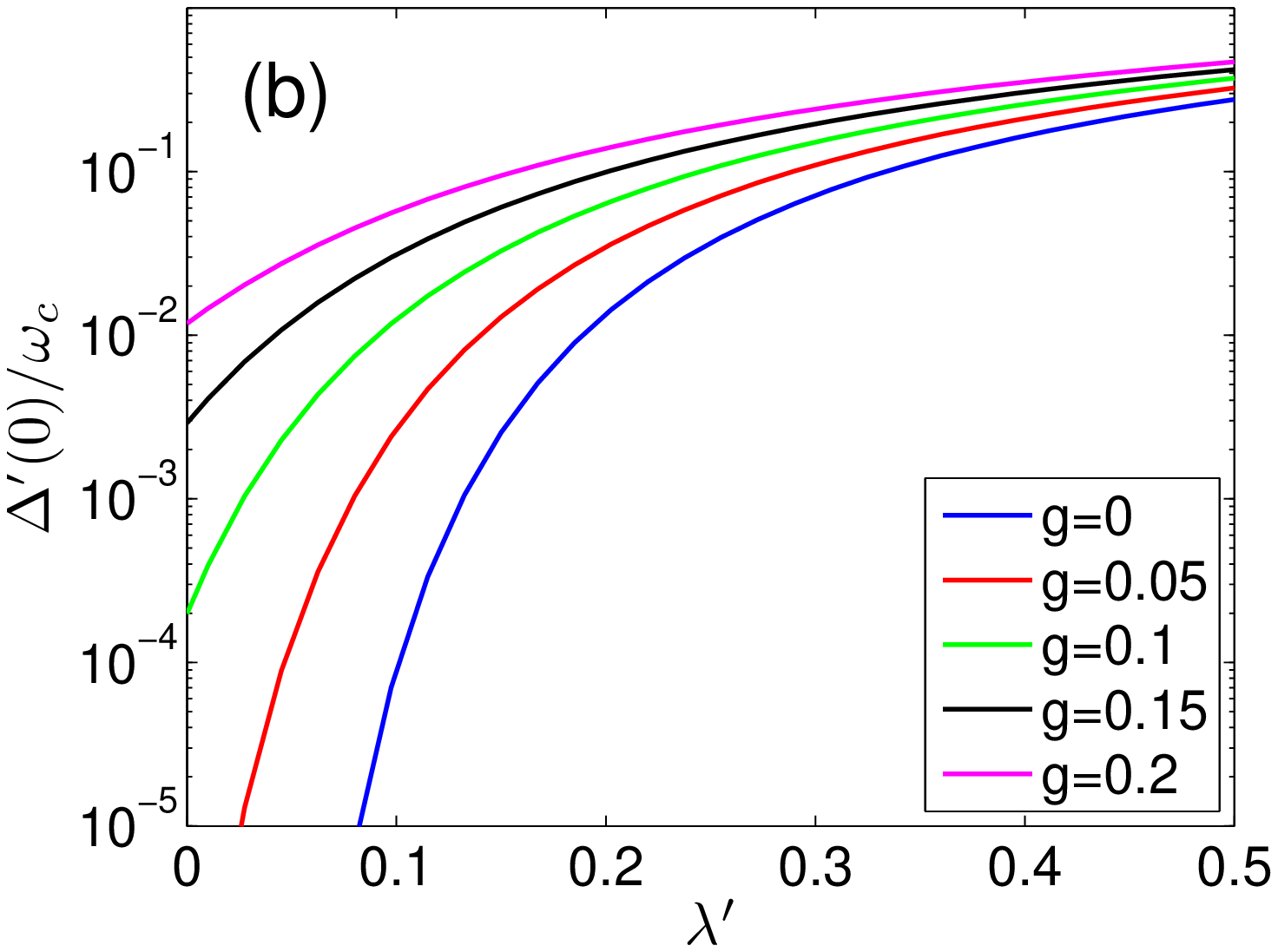}
\includegraphics[width=1.65in]{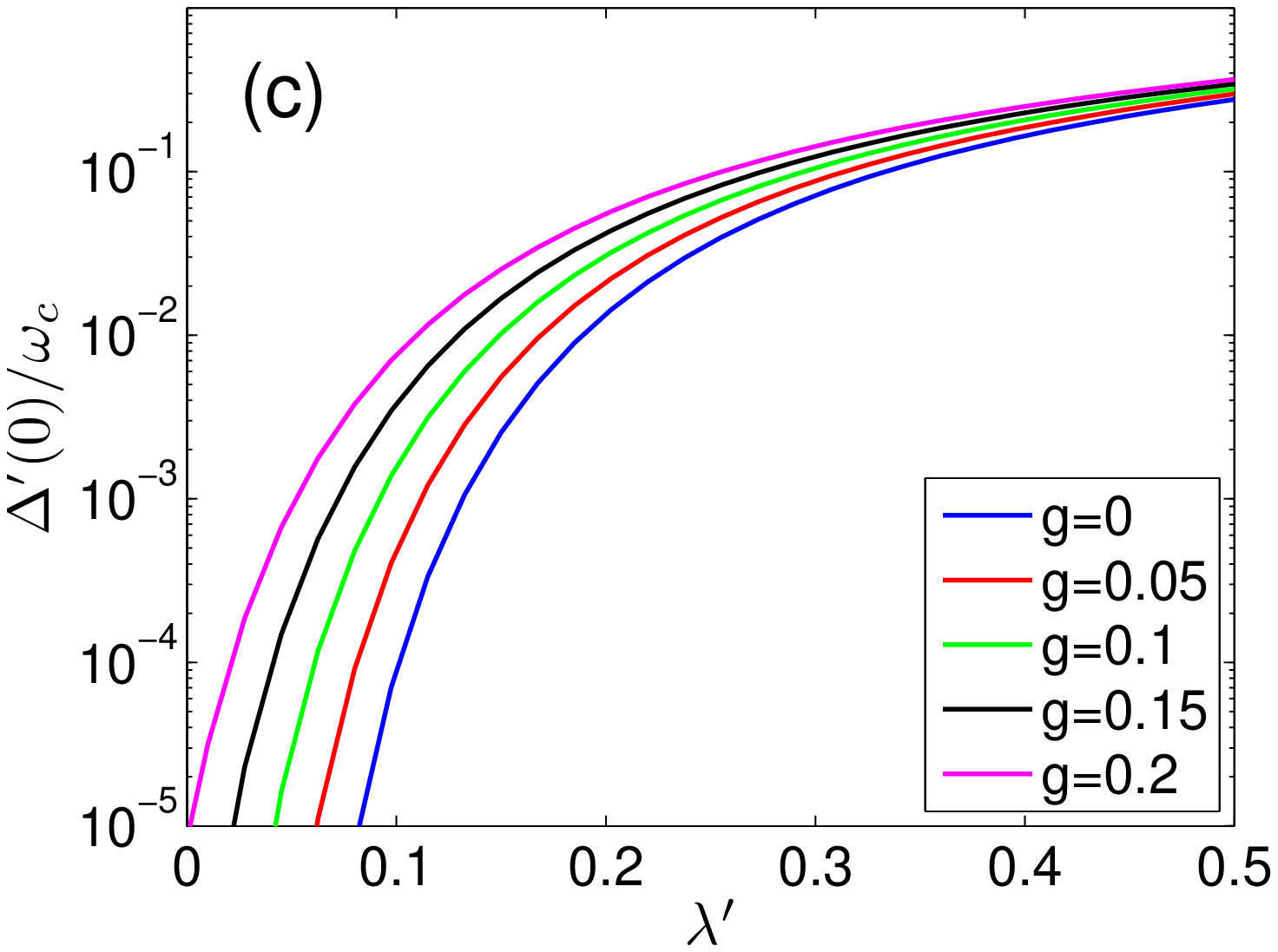}
\includegraphics[width=1.65in]{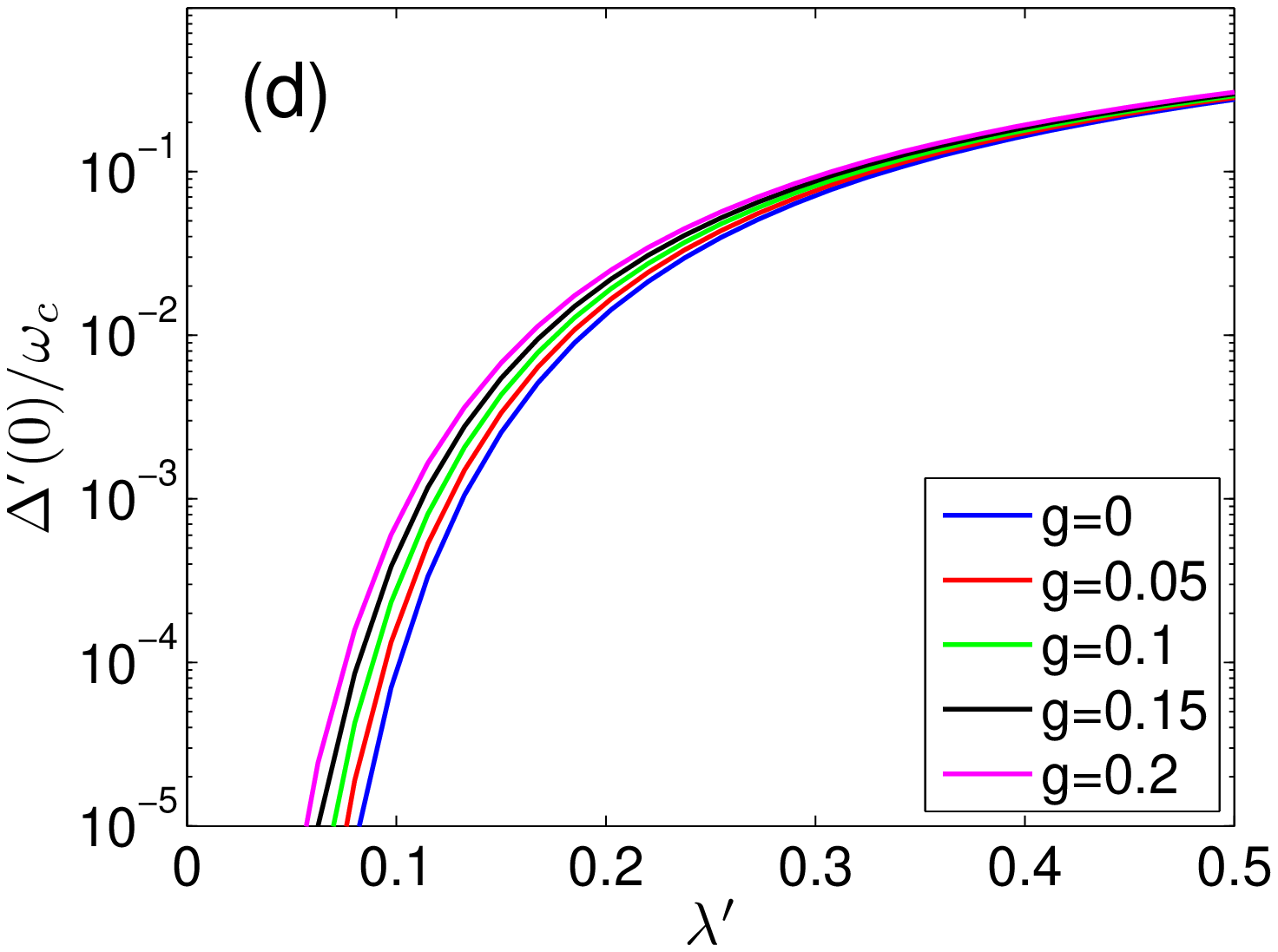}
\caption{Dependence of zero-energy SC gap on tuning parameter $r'$.
The energy scale is given by $\omega_{c}=\mu_{F}$. (a) $r'=0$; (b)
$r'=1$; (c) $r'=10$; (d) $r'=100$. The re-scaled parameter $r' =
e^{2/3}r/(\gamma\omega_{c})^{2/3}$ represents an effective mass of
nematic order parameter. The QCP is located at $r'=0$.
\label{Fig:Gap1DApproBosonMass}}
\end{figure}

To obtain a quantitatively precise relation between the Landau
damping rate and SC gap at $T = 0$, the coupled DS equations in
(\ref{Eq:GeneralEqsZeroT}) cannot be linearized. The linearizing
approximation is valid only in the close vicinity of $T_c$.
Moreover, it is not appropriate to first perturbatively compute
$A_{1}(\varepsilon,p)$ and then to use the perturbative result to solve
the gap equation, because this would miss the important suppressing
effect of SC gap on the Landau damping. In this work, we have solved
the equations in (\ref{Eq:GeneralEqsZeroT}) in a self-consistent and
entirely unbiased way, and determine $A_{1}(\varepsilon,p)$ and
$\Delta(\varepsilon,p)$ simultaneously. To make numerical calculation
simpler, we temporarily ignore the dependence of these function on
momenta, and compute $A_{1}(\varepsilon)$ and $\Delta(\varepsilon)$. The
influence of momentum-dependence will be examined below.

Currently, we concentrate on the nematic QCP and the disordered
phase, corresponding to the parameter range $r \geq 0$. In the
region with $r < 0$, the SC and nematic orders are expected to
coexist. These two orders might compete, which makes theoretical
treatment more involved than the case of $r \geq 0$. This problem
will be considered in future work.

\begin{figure}[htbp]
\includegraphics[width=1.65in]{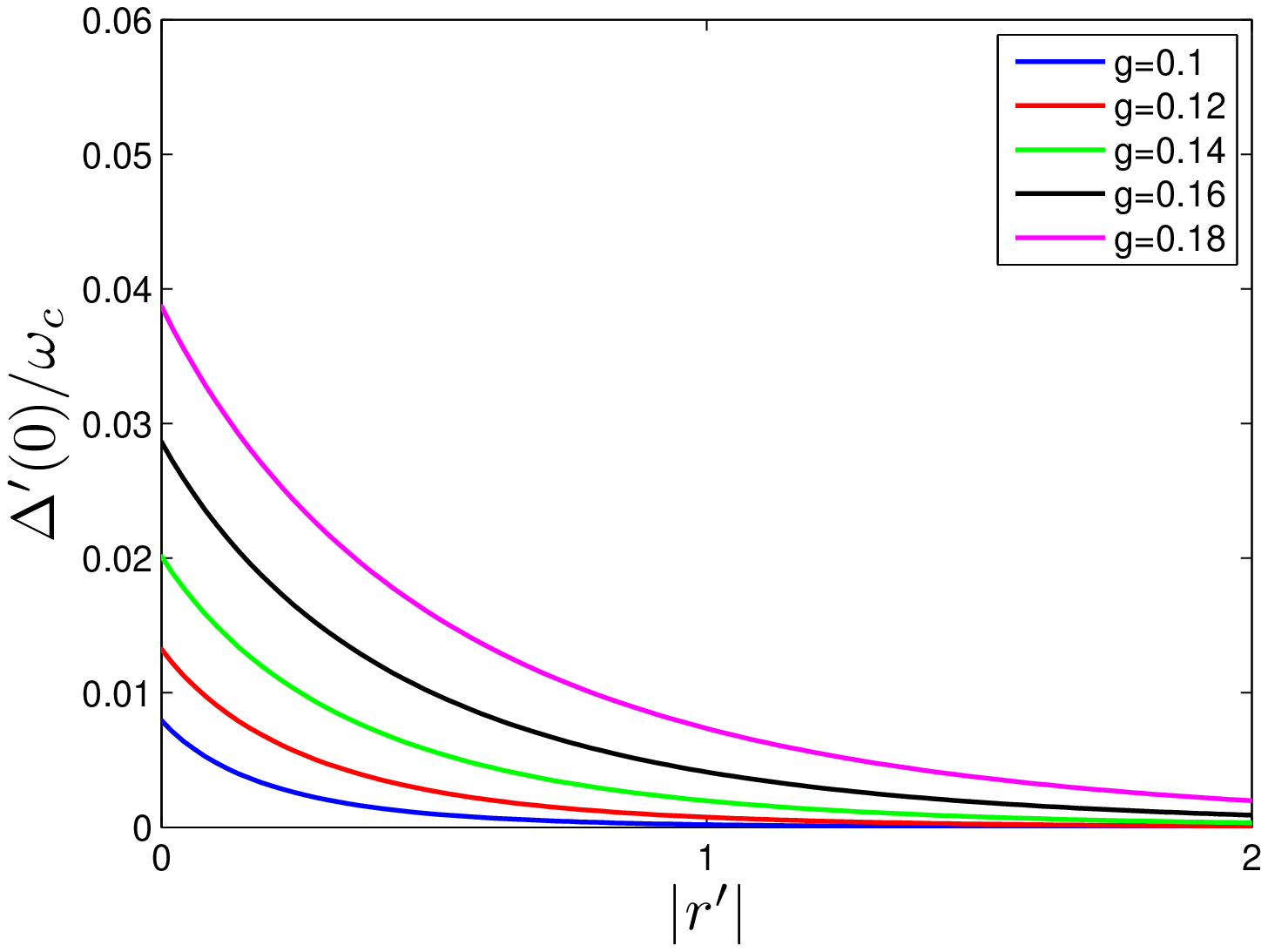}
\includegraphics[width=1.65in]{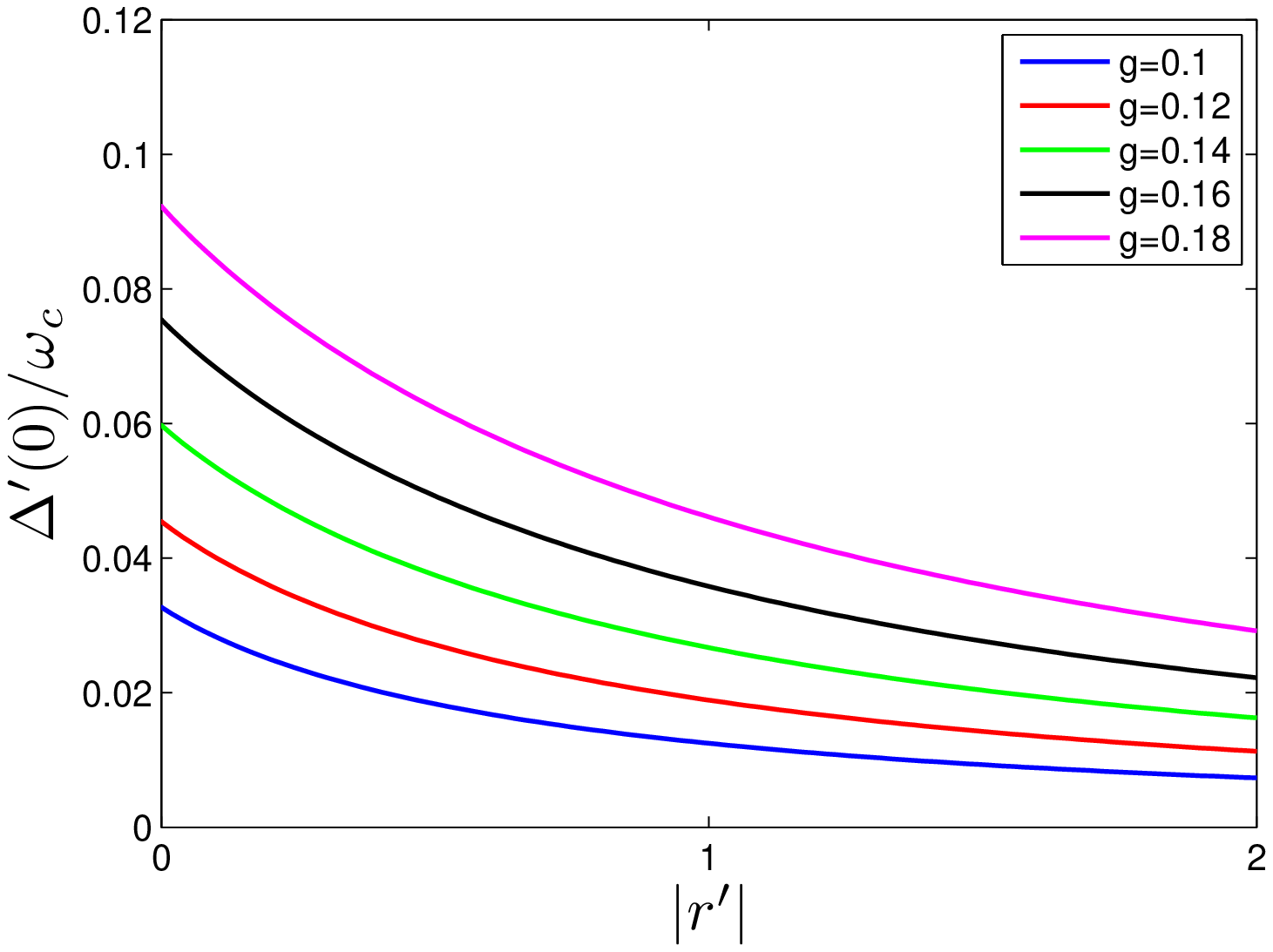}
\caption{Scaled zero-energy SC gap $\Delta'(\omega) =
\Delta(\omega)/A_{1}(\omega)$ is maximized at $r=0$ and is strongly
suppressed by growing $r$. We take $\lambda=0$ in (a) and
$\lambda=0.1$ in (b). The presence of a small $\lambda$ greatly
amplifies the gap size.\label{Fig:GapDome}}
\end{figure}
\begin{figure}[htbp]
\center
\includegraphics[width=1.65in]{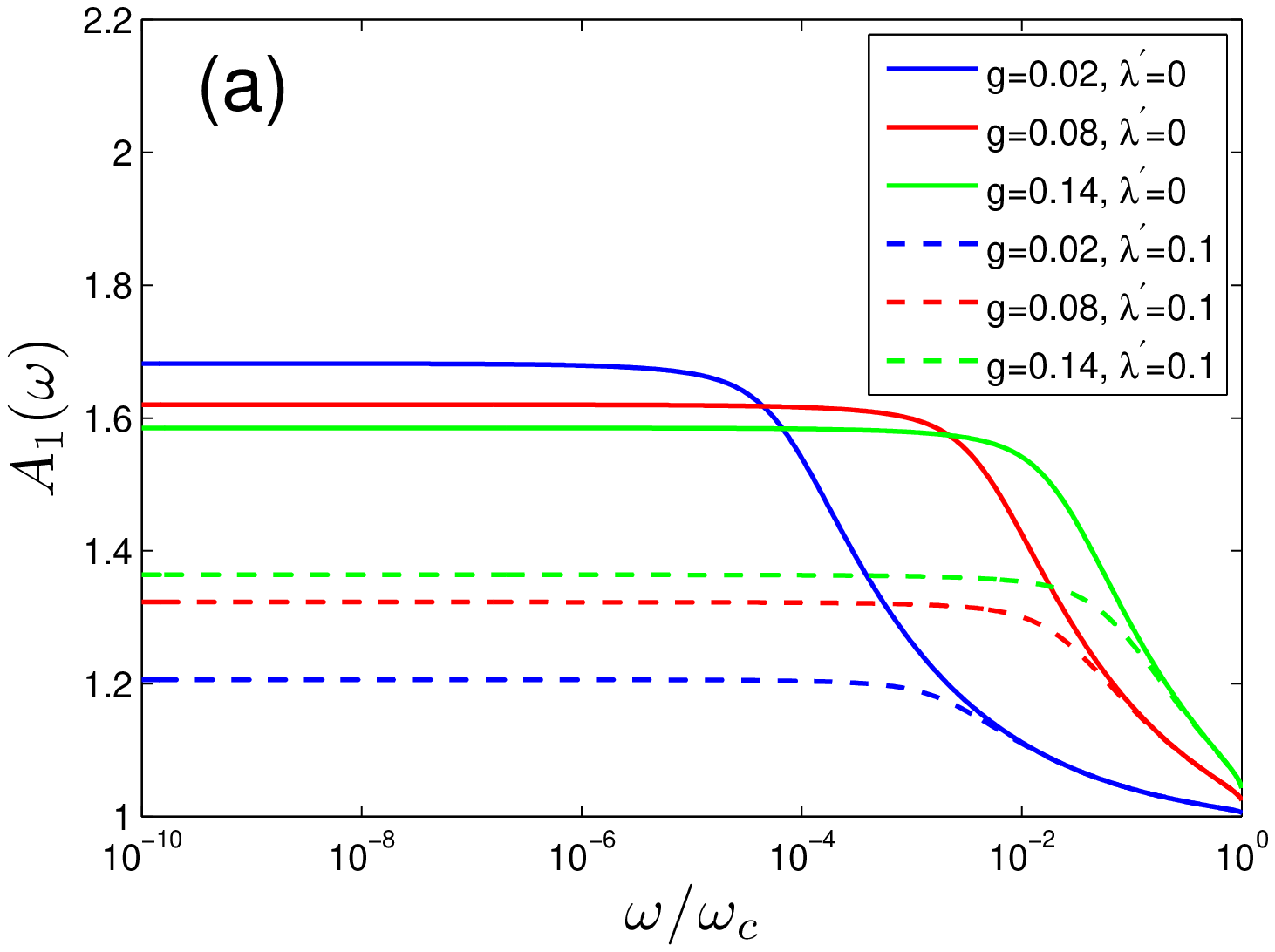}
\includegraphics[width=1.65in]{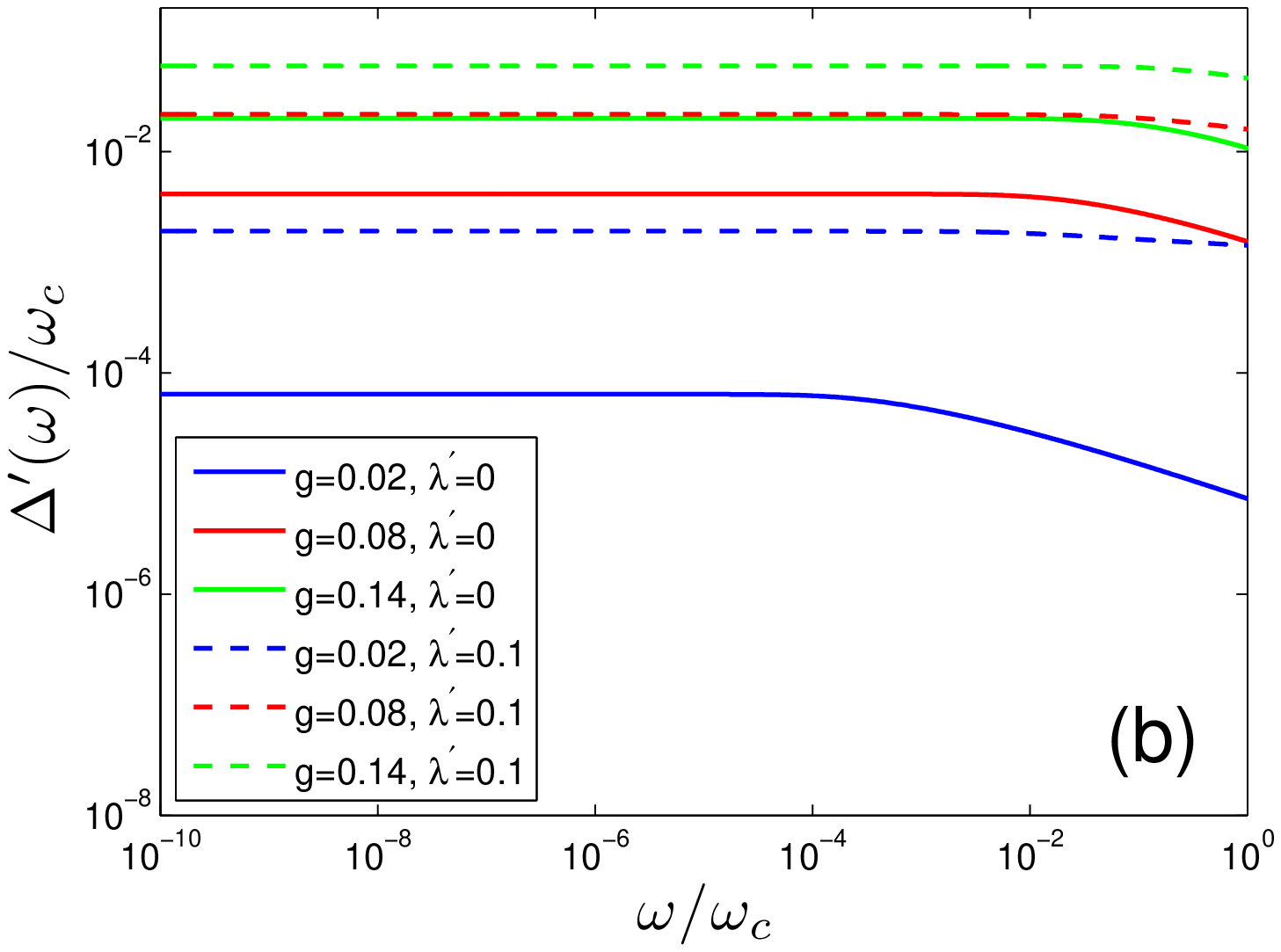}
\caption{Dependence of $A_{1}(\omega)$ and $\Delta(\omega)$ on
$\omega$ are shown in (a) and (b), respectively.
\label{Fig:AFunEGapE}}
\end{figure}

\subsection{Superconducting dome}

In Fig.~\ref{Fig:Gap1DApproBosonMass}, we present the zero-energy SC
gap obtained at various values of $g$, $\lambda'$, and $r'$. At
$r'=0$, the SC gap $\Delta(0)$ obtained at $\lambda'=0.1$ and
$g=0.2$ is roughly one hundred times larger than that obtained at
$\lambda'=0.1$ and $g=0$. As $g$ grows, the gap $\Delta(0)$ further
increases. At $r'=1$, the enhancement is still dramatic. As $r'$
continues growing, the enhancement is rapidly weakened, and finally
nearly disappears when $r'$ is large enough. It is therefore clear
that the nematic-induced enhancement of SC gap is most significant
at the QCP.

The quantum nematic fluctuation itself can trigger Cooper pairing
even though there is no net attraction induced by other scenarios,
consistent with previous works \cite{Metlitski15, Lederer15}. As can
be seen from Fig.~\ref{Fig:Gap1DApproBosonMass}, a finite SC gap is
opened at $\lambda' = 0$, and $\Delta(0)$ is an increasing function
of $g$, the Yukawa coupling. This gap is strongly peaked at the QCP
with $r' = 0$, and decreases rapidly as $r'$ grows. In case a net
attraction with strength $\lambda'$ already develops, presumably due
to commonly existing phonons, the SC gap is significantly enhanced.
Normally, $\lambda'$ only has weak dependence on $r'$, thus the gap
is still peaked at the QCP. All these results are summarized in
Fig.~\ref{Fig:GapDome}. An apparent conclusion is that there emerges
a dome-shaped boundary of the SC phase with the maximal gap
appearing at nematic QCP.

The energy dependence of $A_1(\omega)$ and $\Delta(\omega)$ is shown
in Fig.~\ref{Fig:AFunEGapE}. In the low-energy region, corresponding
to small values of $\omega$, $A_1(\omega)$ is nearly a constant. The
reason for this feature is that the SC gap substantially reduces the
space of final states into which the fermions are scattered by the
nematic fluctuation. The SC gap can be considered as an infrared
cutoff, and the original singular increasing of $A_{1}$ is prevented
in the energy scale lower than the SC gap. It is easy to observe
from Fig.~\ref{Fig:AFunEGapE} that $A_1(\omega)$ is dramatically
suppressed when $\omega/\omega_c$ exceeds certain threshold,
implying the emergence of strong Landau damping effect and,
accordingly, unusual NFL behavior. Above $T_c$, the gap is closed
due to thermal fluctuations, and the system enters from SC phase
into a finite-$T$ NFL phase. If $r'$ takes an intermediate value,
the system is in a mixed FL/NFL regime: the NL and NFL behaviors
show up at different energy scales. All these complicated properties
can be quantitatively reproduced from the self-consistent solutions
of DS equations.

\subsection{Importance of vertex correction}

\begin{figure}[htbp]
\center
\includegraphics[width=1.65in]{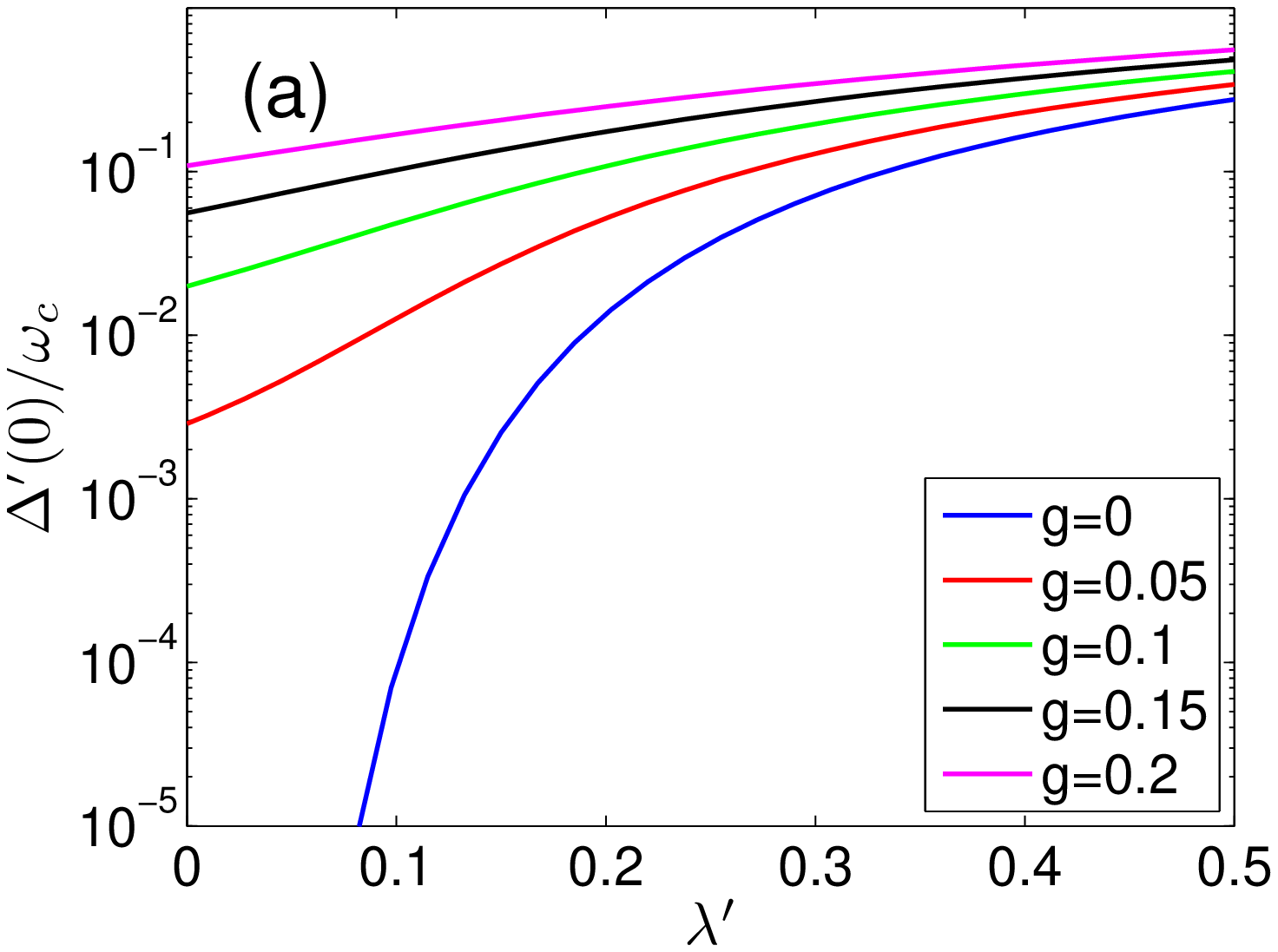}
\includegraphics[width=1.65in]{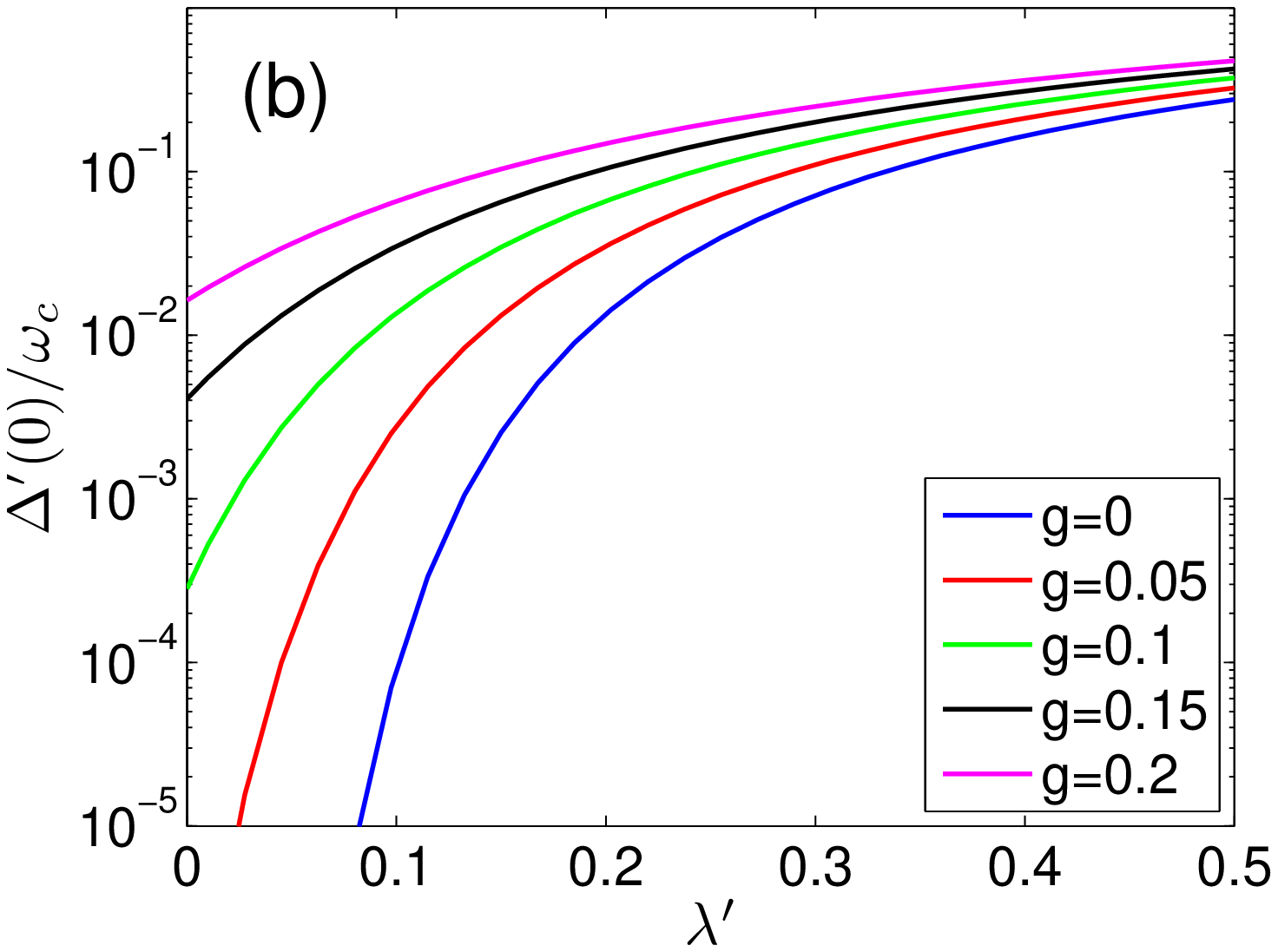}
\caption{Zero-energy SC gap $\Delta(0)$ obtained after including
vertex correction: (a) $r'=0$; (b) $r'=1$. The results for $r'=10$
and $r'=100$ are nearly the same as those given in
Fig.~\ref{Fig:Gap1DApproBosonMass}(c) and
Fig.~\ref{Fig:Gap1DApproBosonMass}(d), and thus are not shown here.
\label{Fig:GapVertexCorrection}}
\end{figure}

We now consider the impact of the vertex correction to Yukawa
coupling between fermion and nematic order. The DS equations are
solved with and without vertex correction respectively, with the
results being given in Fig.~\ref{Fig:GapVertexCorrection}. Comparing
Fig.~\ref{Fig:GapVertexCorrection} to
Fig.~\ref{Fig:Gap1DApproBosonMass}, we find that including the
vertex correction does not change the qualitative results obtained
by adopting the bare vertex. However, the vertex correction leads to
considerable enhancement of superconductivity. If the vertex
correction is ignored, the magnitude of SC gap would be
underestimated. Our results indicate that the vertex correction is
important only at small values of $r'$, namely in the close vicinity
of nematic QCP where the Yukawa coupling is singular. If the system
is far from the nematic QCP, it is valid to ignore the vertex
correction, and the SC transition could be described by the
BCS-Eliashberg method.

\subsection{Including gap in the polarization}

The SC gap size replies sensitively on the effective strength of
Yukawa coupling. The free propagator of nematic fluctuation is
$\propto q^{-2}$ at QCP. Such interaction is effectively
long-ranged. The collective particle-hole excitations weakens such
singular interaction, which is reflected by the polarization
$\Pi(\Omega,\mathbf{q})$. In the above calculations, we have used
the expression of $D(\Omega,\mathbf{q})$ given by
Eq.~(\ref{Eq:BosonPropagatorA}). The feedback effect of SC gap on
$D(\Omega,\mathbf{q})$ needs to be carefully examined. Intuitively,
the SC gap suppresses the low-energy DOS of fermions, which is
expected to weaken the screening effect and increase the effective
strength of nematic fluctuation.

Now assume a finite SC gap $\Delta(\omega)$ is generated. Including
$\Delta$ into the polarization function yields
\begin{widetext}
\begin{eqnarray}
\Pi(\Omega,\mathbf{q}) = 2N\sum_{s=\pm1}\int\frac{d\omega}{2\pi}
\int\frac{d^2\mathbf{k}}{(2\pi)^{2}}\frac{-A_{1}(\omega)A_{1}(\omega+\Omega)
\omega(\omega+\Omega) + \xi_{\mathbf{k}}^{s}
\xi_{\mathbf{k+\mathbf{q}}}^{s}-\Delta(\omega)
\Delta(\omega+\Omega)}{\left[A_{1}^{2}(\omega)\omega^{2} +
\left(\xi_{\mathbf{k}}^{s}\right)^{2} +
\Delta^{2}(\omega)\right]\left[A_{1}^{2}(\omega+\Omega)
\left(\omega+\Omega\right)^{2} +
\left(\xi_{\mathbf{k}+\mathbf{q}}^{s}\right)^{2} +
\Delta^{2}(\omega+\Omega)\right]}.
\end{eqnarray}
\end{widetext}
In principle, the polarization should also be coupled
self-consistently to the DS equations for $A_1$ and $\Delta$. This
is in practice difficult to accomplish. Here, our strategy is to
obtain an approximate analytical expression for the polarization.
According to Appendix B, the exact polarization
$\Pi(\Omega,\mathbf{q})$ can be perfectly replaced by the following
simple function
\begin{eqnarray}
\Pi(\Omega,\mathbf{q}) = N\gamma\frac{|\Omega|}{|q_{y}|}
\frac{|\Omega|}{|\Omega|+2.5\Delta'(0)},
\end{eqnarray}
which reduces to Eq.~(\ref{Eq:BosonPropagatorA}) in the limit
$\Delta'(0) = 0$. It is easy to observe that the above polarization
is considerably smaller than Eq.~(\ref{Eq:PolarizationA}). This
indicates that the effective nematic fluctuation is strengthened
once the SC gap is included in the polarization, which, as just
mentioned, is due to the gap-induced suppression of fermion DOS.

\begin{figure}[htbp]
\center
\includegraphics[width=1.8in]{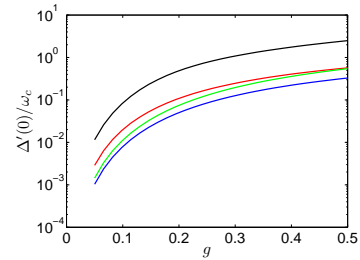}
\caption{Scaled gap obtained under various approximations. Blue:
Vertex and feedback effect are neglected. Red: Vertex is included.
Green: Feedback effect is included. Black: Both vertex and feedback
effect are included.\label{Fig:GapOnlyNematicDiffApp}}
\end{figure}

The SC gap obtained by utilizing different approximations are
presented in Fig.~\ref{Fig:GapOnlyNematicDiffApp}. It clearly shows
that the magnitude of SC gap is visibly enhanced once the feedback
effect of SC gap on the polarization is included.

\subsection{Strong enhancement of superconductivity}

We now analyze the interplay of two different pairing mechanisms.
After long-term exploration, it has become clear that one single
pairing interaction can hardly produce the observed high $T_c$ of
some cuprate and iron-based superconductors. Recently, there is a
growing interest in the study of the cooperative effect of two
distinct pairing interactions \cite{Gorkov16, LiZiXiang16,
Kangjian16, Labat17}. However, the physical influence of such
cooperation remains unclear due to the lack of a well-controlled
framework to properly deal with the interplay of two pairing
interactions.

We will apply the DS equation approach to compute the SC gap induced
by the interplay between nematic fluctuation and additional
short-ranged BCS coupling. Before carrying out calculations, it is
useful to first make a qualitative analysis. As demonstrated in the
last three subsections, the Yukawa coupling between fermions and
nematic fluctuation is strongest at the QCP, and can lead to the
largest SC gap. The Yukawa coupling could be made stronger if the
polarization $\Pi(\Omega,\mathbf{q})$ is reduced. Now imagine a
finite SC gap is already opened by weak BCS coupling, which might be
mediated by the exchange of ordinary phonons. This gap can lower the
low-energy fermion DOS. When the nematic fluctuation is introduced
to the system, the effective strength of Yukawa coupling will be
larger than the case in which no additional BCS coupling exists.
Such scenario is similar to the feedback effect discussed in the
last subsection. In actual materials, the interplay between quantum
nematic fluctuation and electron-phonon interaction could combine to
generate a greatly enhanced superconductivity that can never be
realized by one single pairing interaction.


Here, we would like to mention that an analogous phenomenon occurs
in graphene. It is known that the Coulomb interaction remains
long-ranged despite the presence of dynamical screening due to
particle-hole excitations in graphene. When the Coulomb interaction
is sufficiently strong, it can drive an excitonic pairing and open a
finite dynamical gap, which turns the semimetal into an excitonic
insulator \cite{Khveshchenko01, Gorbar02, Liu09, WangLiu12,
Carrington16}. For strictly gapless graphene, a dynamical excitonic
gap is generated only when the Coulomb interaction strength $\alpha$
is larger than a critical value $\alpha_c$ \cite{Gorbar02, Liu09,
WangLiu12, Carrington16}. Remarkably, if a bare gap is already
opened for some reason, $\alpha_c$ is reduced to an arbitrarily
small value \cite{ZhangLiu11} and the dynamical gap is also
drastically amplified to a much larger value. The strong enhancement
of excitonic pairing originates from the fact that the bare gap
weakens the dynamical screening and increase the effective strength
of Coulomb interaction.

\begin{figure}[htbp]
\center
\includegraphics[width=1.65in]{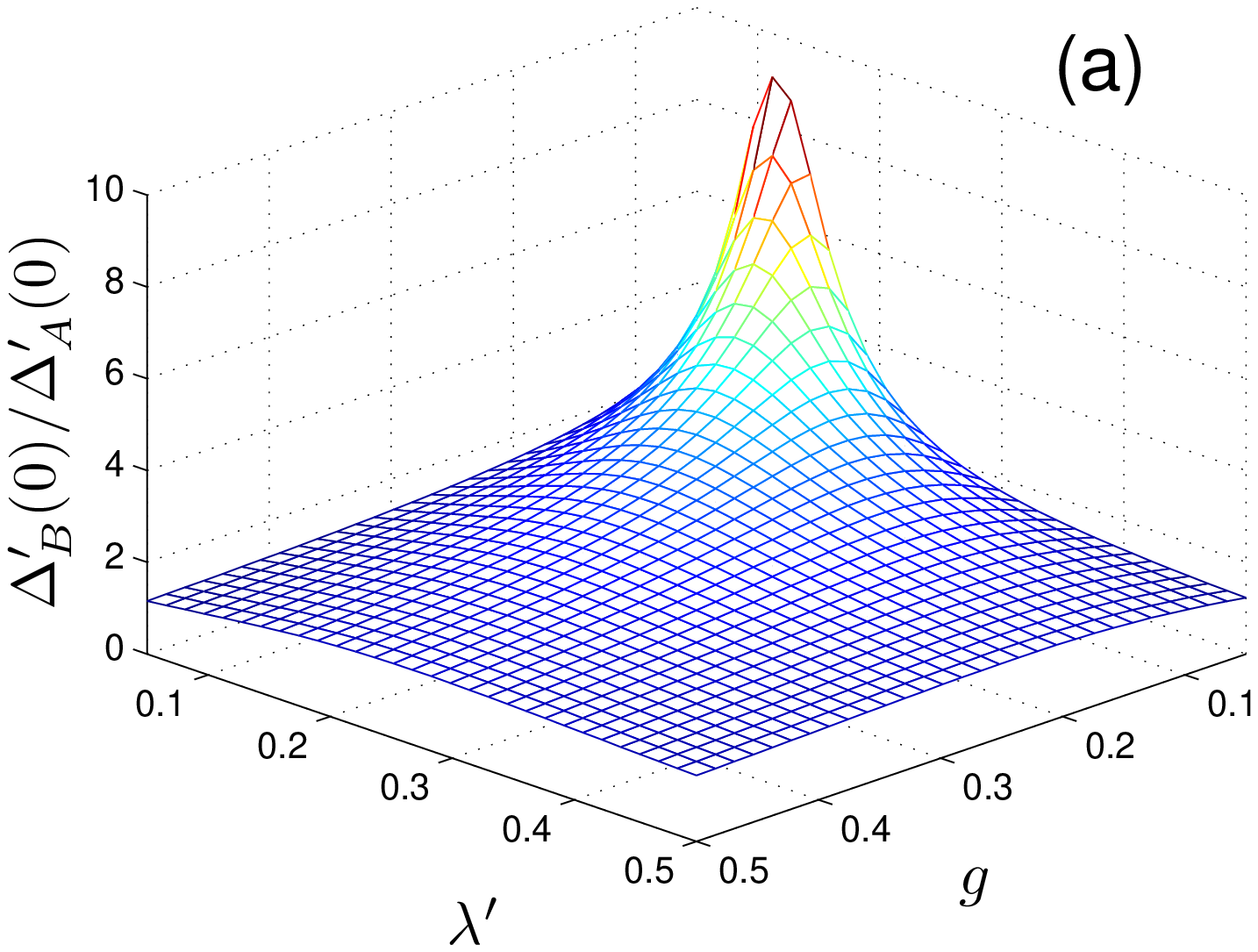}
\includegraphics[width=1.65in]{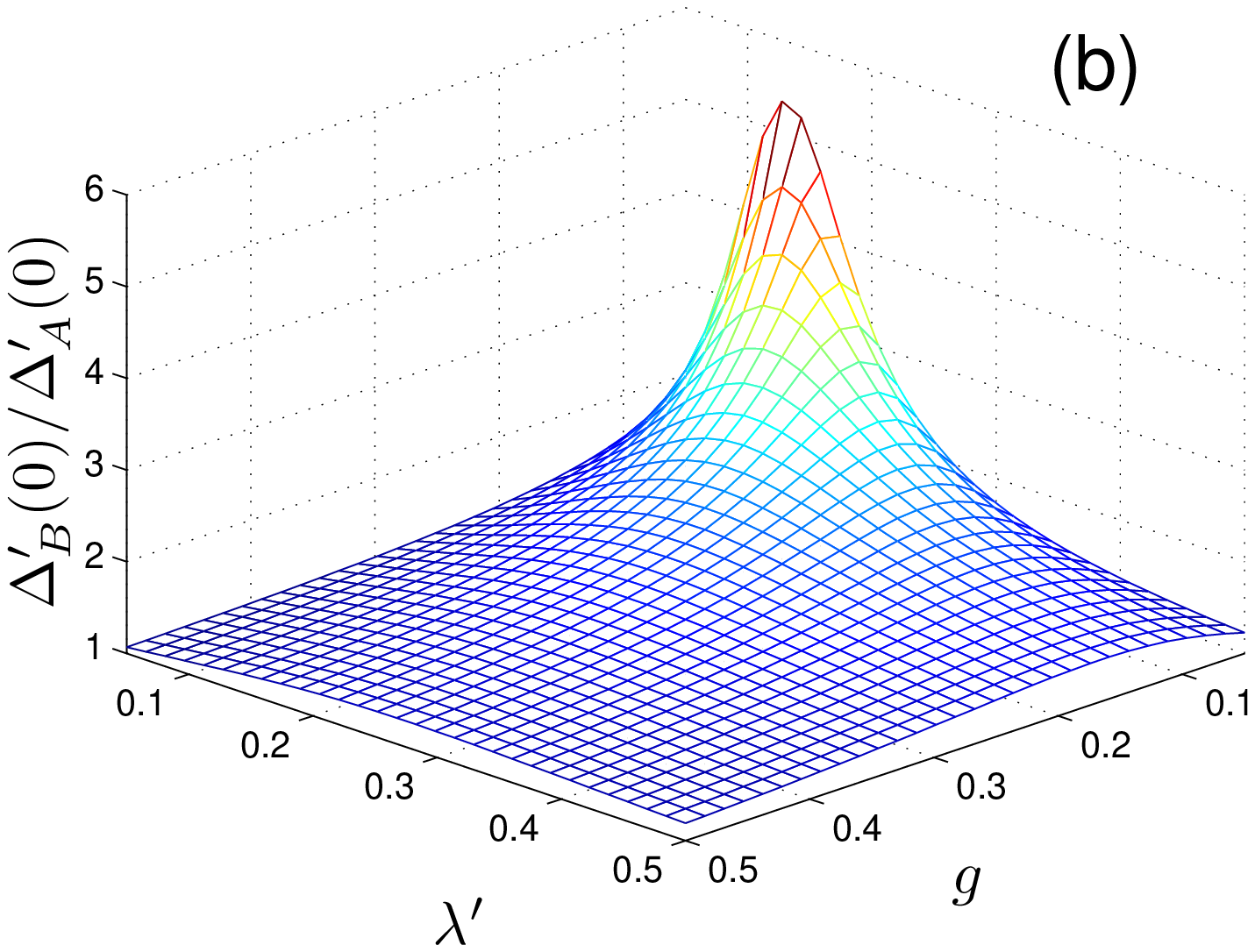}
\includegraphics[width=1.65in]{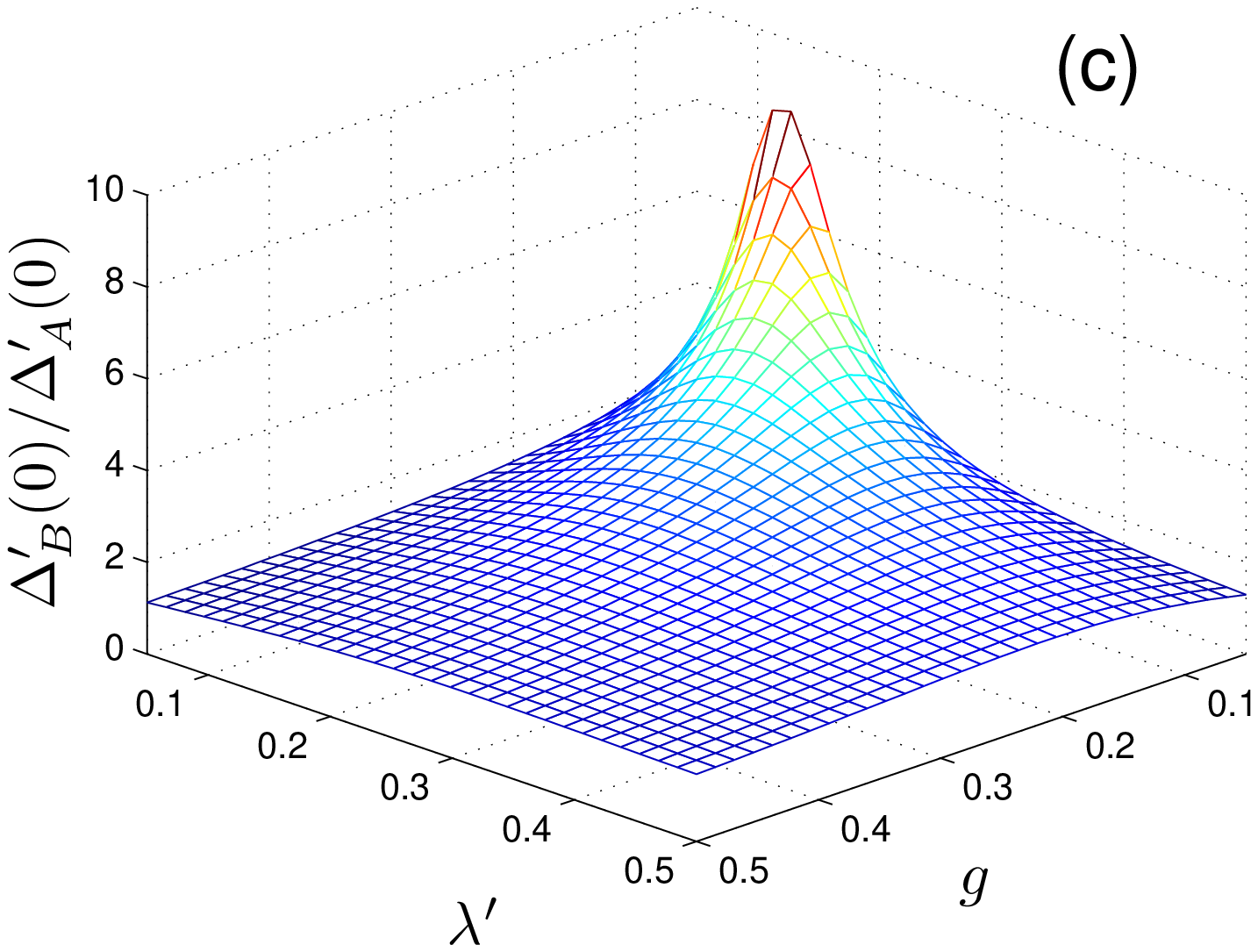}
\includegraphics[width=1.65in]{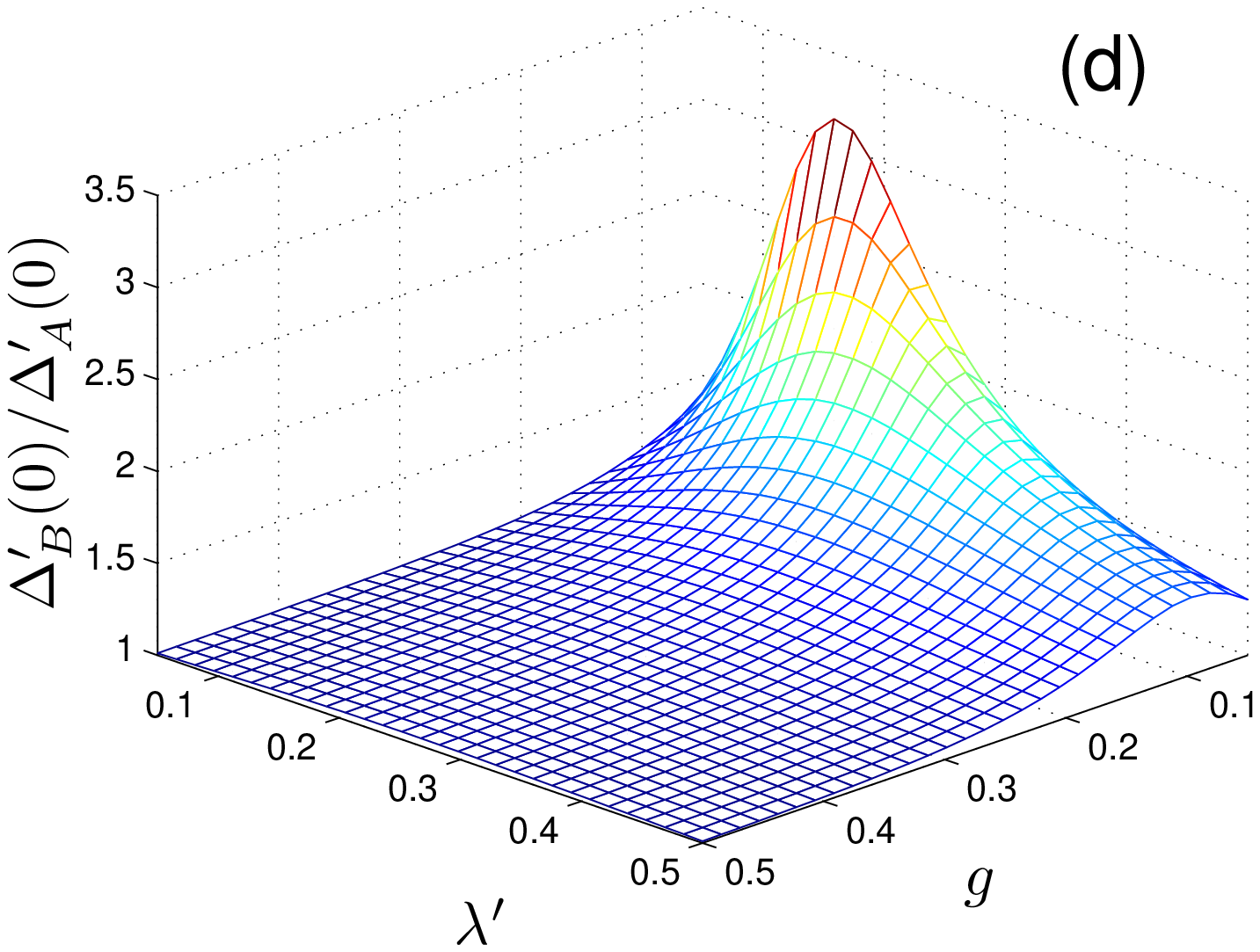}
\caption{Ratio $\Delta'_{B}(0)/\Delta'_{A}(0)$, where
$\Delta_{B}'(0)=\left.\Delta'(0)\right|_{\lambda',g}$ and
$\Delta_{A}'(0)=\left.\Delta'(0)\right|_{\lambda',g=0} +
\left.\Delta'(0)\right|_{\lambda'=0,g}$. (a) Vertex correction and
feedback effect of finite SC gap on the polarization are neglected;
(b) Vertex correction is included; (c) Feedback effect is included;
(d) Both vertex correction and feedback effect are considered.
\label{Fig:RatioInterplay}}
\end{figure}

The significant gap enhancement is firmly based on an important fact
that the superposition of the gaps produced by two pairing
interactions is highly nonlinear. Such nonlinear superposition plays
an important role even when the feedback of SC gap to the Yukawa
coupling is neglected. To precisely evaluate the total SC gap, the
quantum nematic fluctuation and short-ranged BCS coupling should be
treated in a self-consistent way. The DS equation offers a perfect
framework for this study. After solving the DS equations, we find
that the total gap can be much larger than the direct sum of the gap
$\Delta_n$ generated by nematic fluctuation and the gap $\Delta_p$
generated by additional BCS coupling. We present the ratio
$\Delta_{B}(0)/\Delta_{A}$ in Fig.~\ref{Fig:RatioInterplay}, where
$\Delta_{B}(0)$ denotes the total gap and $\Delta_{A}(0) = \Delta_n
+ \Delta_p$. We see that the ratio $\Delta_{B}/\Delta_{A}$ ranges
from unity to 10, depending on the values of model parameters and
the approximation adopted to do the calculation. This ratio is
computed under four different approximations, shown in
Fig.~\ref{Fig:RatioInterplay}(a)-(d). Comparing (a) to (d), we find
that the ratio becomes smaller when both the vertex correction and
feedback to polarization are incorporated. However, it is necessary
to emphasize that the total gap $\Delta_{B}(0)$ obtained in the case
of (d) is indeed much larger than that of case (a).

\subsection{Impact of momentum dependence}

In the above analysis, we entirely ignore the momentum dependence of
$A_{1}(\varepsilon,p_y)$ and $\Delta(\varepsilon,,p_y)$. In Appendix
A, we show how to properly incorporate the momentum dependence in
the coupled DS equations. After performing extensive numerical
calculation, we confirm that our results are only slightly modified
when the momentum dependence of $A_{1}(\varepsilon,p_y)$ and
$\Delta(\varepsilon,,p_y)$ is considered.

\section{Summary and Discussion \label{Sec:Summary}}

To summarize, our work presents a quantitative and self-consistent
determination of the Landau damping rate and the $s$-wave SC gap in
a 2D quantum critical NFL metal. This model system has potential
applications to realistic unconventional superconductors, including
cuprates, iron pnictides and FeSe. We demonstrate that the interplay
of nematic fluctuation and a weak net attraction mediated leads to a
significant enhancement of $s$-wave SC gap, which offers an
efficient way to promote superconductivity. Since the magnitude of
zero-$T$ SC gap is directly related to the SC transition
temperature, the gap enhancement could lead to a remarkably
increased $T_c$. This motivates us to conjecture that the observed
high $T_c$ of some cuprate and iron-based superconductors might
originate from the mutual promotion of two distinct pairing
mechanisms. Moreover, the total SC gap reaches its maximal value at
the nematic QCP and is strongly suppressed as the system is tuned
away from the QCP, hence a dome-shaped curve of $T_c$ could be
naturally produced, which appears to be in general agreement with
experiments.

Our approach can be regarded as an extension of the BCS-Eliashberg
theory to quantum critical metals in which the Landau damping is
strong enough to invalidate the FL theory. Depending on the value of
tuning parameter $r$, the non-SC system might stay in NFL regime, FL
regime, or mixed FL/NFL regime that displays ordinary FL behavior at
low energies and strong NFL behavior at high energies. Once Cooper
pairing of (in)coherent fermions is realized, the system is in the
SC state at low temperatures. Nevertheless, NFL behavior can still
emerge in the intermediate energy range. Our approach thus provides
a unified framework for the theoretic analysis of Cooper pairing in
FL, NFL, and mixed FL/NFL metals.

The present work is restricted to zero temperature. The next step is
to study the influence of finite temperature, and to accurately
compute $T_c(r)$. This is not an easy problem because the quantum
critical nematic fluctuation might lead to severe infrared
divergence in the DS equations at nonzero temperature
\cite{WangLiuZhang15, WangHuaJia17B}. It is also necessary to
investigate the case in which the nematic order parameter has a
finite mean value. The coexistence of nematic and SC orders might
cause unusual effects that cannot occur in the disordered side of
nematic QCP.

Another interesting future work is to apply the DS equation method
to study the fate of superconductivity in correlated electron
systems close to magnetic quantum phase transition. Such systems
have direct applications to iron-based superconductors
\cite{Chubukov12, Fernandes14, Shibauchi14}. The magnetic order
parameter is more complicated than nematic order parameter
\cite{Chubukov2003, Chubukov2004} and the SC gap induced by magnetic
fluctuation may have a $d$-wave symmetry, which makes DS equation
analysis more involved. In some superconductors, the magnetic and
nematic long-range orders are both important and indeed
intrinsically connected \cite{Fernandes14, Shibauchi17}. Despite
such complications, one can always construct a set of coupled DS
equations for the SC gap function and the renormalization factors,
analyze the structure of the gap, determine the correlation between
NFL behavior and Cooper pairing, and also examine whether the
interplay of distinct pairing mechanisms lead to significantly
enhanced superconductivity.

We would like to thank Jing Wang and Chun-Xu Zhang for very helpful
discussions, and acknowledge the support by the National Natural
Science Foundation of China under Grants 11574285 and 11504379.

\clearpage

\newpage

\appendix

\begin{widetext}

\section{Influence of momentum dependence of }

Here we examine whether the momentum dependence of $A_{1}$ and
$\Delta$ play an important role. It is in principle to solve the
self-consistent equations (\ref{Eq:A12DGeneral}) and
(\ref{Eq:Gap2DGeneral}) numerically. Nevertheless, this is
technically hard and extremely time-consuming. We choose to
factorize the functions $A_{1}(\varepsilon,p_{y})$ and
$\Delta(\varepsilon,p_{y})$ as follows:
\begin{eqnarray}
A_{1}(\varepsilon,p_{y}) &=& A_{1}^{a}(\varepsilon)F_{1}(p_{y}),
\\
\Delta(\varepsilon,p_{y}) &=& \Delta{^a}(\varepsilon)F_{2}(p_{y}),
\end{eqnarray}
where $F_{1}$ and $F_{2}$ satisfy
\begin{eqnarray}
F_{1}(0) &\equiv& 1, \\
F_{2}(0) &\equiv& 1.
\end{eqnarray}
Accordingly, the self-consistent equations can be re-written as
\begin{eqnarray}
A_{1}^{a}(\varepsilon)\varepsilon &=& \varepsilon + \frac{1}{2v}
\int\frac{d\omega}{2\pi}\frac{dk_{y}}{2\pi} \frac{\Gamma
\left(\varepsilon,0;\omega,k_{y}\right) A_{1}^{a}(\omega)
F_{1}(k_{y})\omega}{\sqrt{\left(A_{1}^{a}(\omega)\right)^{2}
F_{1}^{2}\left(k_{y}\right)\omega^{2} +
\left(\Delta^{a}(\omega)\right)^{2}
F_{2}^{2}(k_{y})}}D(\varepsilon-\omega,k_{y}), \\
A_{1}^{a}(\omega_{c})F_{1}(p_{y})\omega_{c} &=& \omega_{c} +
\frac{1}{2v}\int\frac{d\omega}{2\pi}\frac{dk_{y}}{2\pi}
\frac{\Gamma\left(\omega_{c},p_{y};\omega,k_{y}\right)
A_{1}^{a}(\omega)F_{1}(k_{y})\omega}{\sqrt{\left(A_{1}^{a}(\omega)\right)^{2}
F_{1}^{2}\left(k_{y}\right)\omega^{2} +
\left(\Delta^{a}(\omega)\right)^{2}F_{2}^{2}(k_{y})}}
D(\omega_{c}-\omega,p_{y}-k_{y}),
\\
\Delta^{a}(\varepsilon)&=&\frac{\lambda}{2v}\int
\frac{d\omega}{2\pi}\frac{dk_{y}}{2\pi} \frac{\Delta^{a}(\omega)
F_{2}(k_{y})}{\sqrt{\left(A_{1}^{a}(\omega)\right)^{2}
F_{1}^{2}\left(k_{y}\right)\omega^{2} +
\left(\Delta^{a}(\omega)\right)^{2}F_{2}^{2}(k_{y})}}\nonumber
\\
&&+\frac{1}{2v}\int\frac{d\omega}{2\pi}\frac{dk_{y}}{2\pi}
\frac{\Gamma\left(\varepsilon,0;\omega,k_{y}\right)\Delta^{a}(\omega)F_{2}(k_{y})}
{\sqrt{\left(A_{1}^{a}(\omega)\right)^{2}F_{1}^{2}\left(k_{y}\right)\omega^{2}
+\left(\Delta^{a}(\omega)\right)^{2}F_{2}^{2}(k_{y})}}D(\varepsilon-\omega,k_{y}),
\\
\Delta^{a}(\omega_{c})F_{2}(p_{y}) &=& \frac{\lambda}{2v}\int
\frac{d\omega}{2\pi}\frac{dk_{y}}{2\pi}\frac{\Delta^{a}(\omega)
F_{2}(k_{y})}{\sqrt{\left(A_{1}^{a}(\omega)\right)^{2}
F_{1}^{2}\left(k_{y}\right)\omega^{2} +
\left(\Delta^{a}(\omega)\right)^{2}F_{2}^{2}(k_{y})}}\nonumber
\\
&&+\frac{1}{2v}\int\frac{d\omega}{2\pi}\frac{dk_{y}}{2\pi}
\frac{\Gamma\left(\omega_{c},p_{y};\omega,k_{y}\right)
\Delta^{a}(\omega)F_{2}(k_{y})}{\sqrt{\left(A_{1}^{a}(\omega)\right)^{2}
F_{1}^{2}\left(k_{y}\right)\omega^{2}
+\left(\Delta^{a}(\omega)\right)^{2}F_{2}^{2}(k_{y})}}
D(\omega_{c}-\omega,p_{y}-k_{y}).
\end{eqnarray}
By solving these equations, we have confirmed that the momentum
dependence of $A_{1}$ and $\Delta$ can only play a minor role. The
zero-energy gap $\Delta'(0)$ obtained in the presence and absence of
such momentum dependence are nearly the same.

\begin{figure}[htbp]
\center
\includegraphics[width=2.8in]{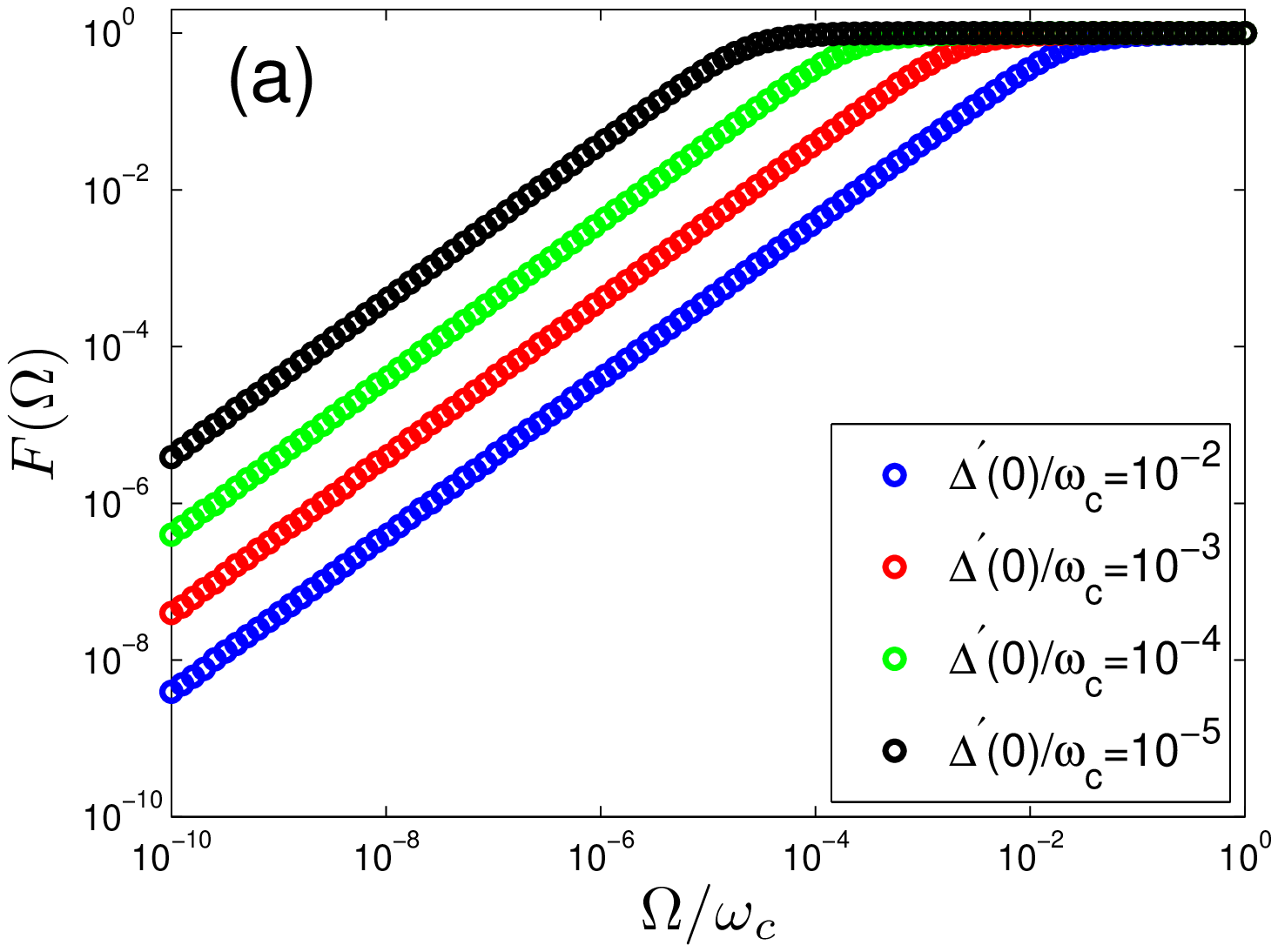}
\includegraphics[width=2.8in]{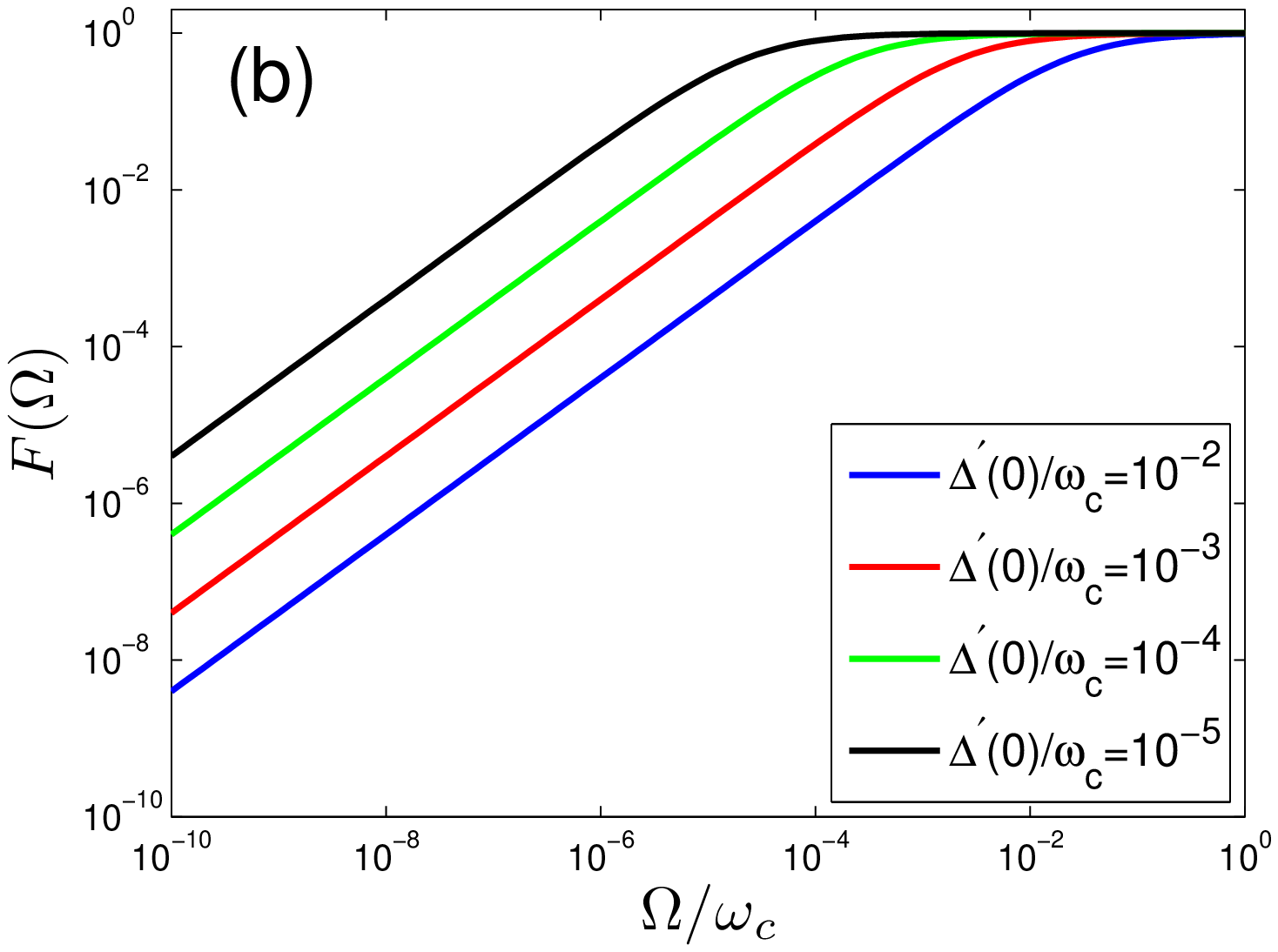}
\caption{(a) Numerical results for $F(|\Omega|)$ with different
values $\Delta'(0)$; (b) The analytical expression
$F(|\Omega|)=\frac{|\Omega|}{|\Omega|+2.5\Delta'(0)}$ with different
values of $\Delta'(0)$. \label{Fig:FFunNumFit}}
\end{figure}

\section{Including gap in the polarization}

Including the feedback effect of finite SC gap, the polarization
takes the following form
\begin{eqnarray}
\Pi(\Omega,\mathbf{q}) &=& 2N\sum_{s=\pm1}\int\frac{d\omega}{2\pi}
\int\frac{d^2\mathbf{k}}{(2\pi)^{2}} \nonumber
\\
&& \times \frac{-A_{1}(\omega)A_{1}(\omega+\Omega)
\omega(\omega+\Omega) + \xi_{\mathbf{k}}^{s}
\xi_{\mathbf{k+\mathbf{q}}}^{s}-\Delta(\omega)\Delta(\omega+\Omega)}
{\left[A_{1}^{2}(\omega)\omega^{2}+\left(\xi_{\mathbf{k}}^{s}\right)^{2}
+ \Delta^{2}(\omega)\right]\left[A_{1}^{2}(\omega+\Omega)
\left(\omega+\Omega\right)^{2}+\left(\xi_{\mathbf{k}+\mathbf{q}}^{s}\right)^{2}
+ \Delta^{2}(\omega+\Omega)\right]}.
\end{eqnarray}
Performing the integration of momentum, we obtain
\begin{eqnarray}
\Pi(\Omega,\mathbf{q}) &=& \frac{N\gamma}{2|q_{y}|}\int d\omega
\frac{-A_{1}(\omega)A_{1}(\omega+\Omega) \omega(\omega+\Omega) -
\Delta(\omega)\Delta(\omega+\Omega)} {\sqrt{A_{1}^{2}(\omega)
\omega^{2}+\Delta^{2}(\omega)}\sqrt{A_{1}^{2}(\omega+\Omega)
\left(\omega+\Omega\right)^{2}+\Delta^{2}(\omega+\Omega)}}.
\end{eqnarray}
In the limit of $\Omega=0$ and $\Delta=0$, the polarization is
simplified to
\begin{eqnarray}
\Pi(\Omega=0,\mathbf{q},\Delta=0) = N\gamma\frac{1}{2|q_{y}|}\int
d\omega(-1).
\end{eqnarray}
For the polarization to satisfy the condition
$\Pi(\Omega=0,\mathbf{q},\Delta=0)=0$, we employ the redefinition
\begin{eqnarray}
\Pi(\Omega,\mathbf{q})-\Pi(\Omega=0,\mathbf{q},\Delta=0)\rightarrow
\Pi(\Omega,\mathbf{q}),
\end{eqnarray}
and then obtain
\begin{eqnarray}
\Pi(\Omega,\mathbf{q}) &=& N\gamma\frac{1}{2|q_{y}|}\int d\omega
\left[1-\frac{A_{1}(\omega)A_{1}(\omega+\Omega)\omega(\omega+\Omega)
+ \Delta(\omega)\Delta(\omega+\Omega)}{\sqrt{A_{1}^{2}(\omega)
\omega^{2} + \Delta^{2}(\omega)}\sqrt{A_{1}^{2}(\omega+\Omega)
\left(\omega+\Omega\right)^{2}+\Delta^{2}(\omega+\Omega)}}\right].
\end{eqnarray}
It can be further written as
\begin{eqnarray}
\Pi(\Omega,\mathbf{q}) &=& N\gamma\frac{|\Omega|}{|q_{y}|}
F(|\Omega|),
\end{eqnarray}
where
\begin{eqnarray}
F(|\Omega|) &=& \frac{1}{|\Omega|}\int_{0}^{+\infty} d\omega
\left[1-\frac{A_{1}(\omega)A_{1}(\omega+|\Omega|)\omega(\omega+|\Omega|)
+ \Delta(\omega)\Delta(\omega+|\Omega|)}{\sqrt{A_{1}^{2}(\omega)
\omega^{2} + \Delta^{2}(\omega)}\sqrt{A_{1}^{2}(\omega +
|\Omega|)(\omega + |\Omega|)^{2} + \Delta^{2}(\omega +
|\Omega|)}}\right] \nonumber \\
&&+\frac{1}{2|\Omega|}\int_{0}^{|\Omega|} d\omega
\left[1-\frac{-A_{1}(\omega)A_{1}(\omega-|\Omega|)\omega(-\omega+|\Omega|)
+\Delta(\omega)\Delta(-\omega+|\Omega|)}{\sqrt{A_{1}^{2}(\omega)
\omega^{2}+\Delta^{2}(\omega)}\sqrt{A_{1}^{2}(\omega-|\Omega|)
\left(\omega-|\Omega|\right)^{2} +
\Delta^{2}(\omega-|\Omega|)}}\right]. \nonumber
\end{eqnarray}
Under the approximation $A_{1}(\omega) \approx A_{1}(0)$ and
$\Delta(\omega)\approx \Delta(0)$, we get
\begin{eqnarray}
F(|\Omega|) &=& \frac{1}{|\Omega|}\int_{0}^{+\infty}
d\omega\left[1-\frac{\omega(\omega + |\Omega|)+\Delta'^{2}(0)}
{\sqrt{\omega^{2}+\Delta'^{2}(0)}\sqrt{(\omega+|\Omega|)^{2} +
\Delta'^{2}(0)}}\right]\nonumber \\
&&+\frac{1}{2|\Omega|}\int_{0}^{|\Omega|} d\omega \left[1 -
\frac{-\omega(-\omega+|\Omega|)+\Delta'^{2}(0)}{\sqrt{\omega^{2} +
\Delta'^{2}(0)}\sqrt{\left(-\omega+|\Omega|\right)^{2} +
\Delta'^{2}(0)}}\right],
\end{eqnarray}
where $\Delta'(0)=\frac{\Delta(0)}{A_{1}(0)}$. The numerical result
is shown in Fig.~\ref{Fig:FFunNumFit}, which indicates that
$F(|\Omega|)$ can be accurately approximated by the expression
\begin{eqnarray}
F(|\Omega|)=\frac{|\Omega|}{|\Omega|+2.5\Delta'(0)}.
\end{eqnarray}
\end{widetext}

\end{document}